\documentclass[fleqn,10pt]{wlscirep}

\usepackage[utf8]{inputenc}
\usepackage[T1]{fontenc}
\usepackage{bm,hyperref}
\usepackage{amsfonts} 
\usepackage{colortbl}
\usepackage{verbatim}
\usepackage{setspace}
\usepackage{soul}
\title{Multidimensional optical singularities and their applications}

\author[1,$\dagger$]{Soon Wei Daniel Lim}
\author[1,$\dagger$]{Christina M. Spaegele}
\author[1,*]{Federico Capasso}
\affil[1]{Harvard John A. Paulson School of Engineering and Applied Sciences, Harvard University, Cambridge,
MA 02138, USA
}
\affil[*]{e-mail: capasso@seas.harvard.edu}
\affil[$\dagger$]{equal contribution}

\begin{abstract}
Optical singularities, which are positions within an electromagnetic field where certain field parameters become undefined, hold significant potential for applications in areas such as super-resolution microscopy, sensing, and communication. This potential stems from their high field confinement and characteristic rapidly-changing field distributions. Although the systematic characterization of the first singularities dates back many decades, recent advancements in sub-wavelength wavefront control at optical frequencies have led to a renewed interest in the field, and have substantially expanded the range of known optical singularities and singular structures. However, the diversity in descriptions, mathematical formulations, and naming conventions can create confusion and impede accessibility to the field. This review aims to clarify the nomenclature by demonstrating that any singular field can be conceptualized as a collection of a finite set of principal, 'generic' singularities. These singularities are robust against small perturbations due to their topological nature. We underscore that the control over the principal properties of those singularities, namely, their protection against perturbations and their dimension, utilizes a consistent mathematical framework. Additionally, we provide an overview of current design techniques for both stable and approximate singularities and discuss their applications across various disciplines.
\end{abstract}

\begin{document}
\renewcommand{\hbar}{\mathchar'26\mkern-9mu h} 

\newcommand{\eg}{\textit{e.g., }}
\newcommand{\ie}{\textit{i.e., }}

\flushbottom
\maketitle

\thispagestyle{empty}

\tableofcontents

\section{Introduction}
Optical singularities are positions in an optical field at which a certain parameter is singular (\ie undefined). The canonical instances are phase singularities, \ie points of undefined phase in a complex scalar field, which typically manifest as random speckle patterns and are integral to multiple imaging and sensing applications. Notably, the Nobel Prize-winning work of Stefan Hell in Stimulated Emission Depletion (STED) super-resolution microscopy employs the geometry of phase singularities to surpass the fundamental diffraction limit by an order of magnitude \cite{1994_Hell_Wichmann_OptLett,2007_Willig_Hell_NatMethods}. 

The field of singular optics extends far beyond points of singular phase: A quick survey of the literature reveals a multitude of descriptors for optical singularities: d points \cite{1987_Hajnal_PRSLA_II}, lowercase c points \cite{1987_Hajnal_PRSLA_II,2002_Freund_OptComm,2013_Berry_JOpt}, uppercase C points \cite{1987_Nye_Hajnal_PRSL}, S points \cite{2020_Chen_Liu_LaserPhotRev}, L points \cite{2002_Freund_OptComm}, L lines \cite{1987_Hajnal_PRSLA_I,1987_Hajnal_PRSLA_II,2002_Freund_OptComm}, L surfaces \cite{2018_Shvedov_Krolikowski_NewJPhys}, V-lines, V-circles, C-circles \cite{2002_Freund_OptComm}, stars \cite{1983_Nye_PRSLA}, Monstars\cite{1983_Nye_PRSLA}, Lemons \cite{1983_Nye_PRSLA}, $\Sigma$ points \cite{2001_Freund_OptComm,2002_Freund_OptComm}, fractional vortices \cite{2004_Basistiy_Vasnetsov_JOptA,2004_Berry_JOptA_fractional,2004_Leach_Padgett_NewJPhys}, perfect vortices \cite{2013_Ostrovsky_Arrizon_OptLett}, knots and Möbius strips \cite{2001_Berry_Dennis_ProcRoyalSocLonA,2015_Bauer_Leuchs_Science,2019_Tekce_Denz_OptExp}, Skyrmions and Hopfions \cite{2021_Sugic_Dennis_NatComm,2021_Shen_Zheludev_NatComm}, and spin defects \cite{2022_Wang_Fan_Optica}, to name just a few. M. Berry has eloquently distinguished between different terminologies with contradictory definitions for $C$ loci, a type of polarization singularity \cite{2013_Berry_JOpt}. Studying optical singularities is akin to cataloguing a diverse zoo of species. However, at the heart of each optical singularity, only a small set of parameters can be singular: phase, field directions, polarization ellipse parameters, and the polarization state. More complex singular structures can be formed through spatial composition of several fundamental singularities or by representation in a different parameter space, such as in 3D Cartesian space instead of the 2D plane. 

The creation and manipulation of various optical singularities requires the use of devices which can control these field parameters. While phase singularities can be created by judicious superposition of modes or plane waves\cite{1987_Hajnal_PRSLA_II,1992_Allen_Woerdman_PhysRevA}, they are easily created with devices that have direct control over the wavefront, such as phase or amplitude holographic plates \cite{1992_Heckenberg_White_OptLett,1992_Bazhenov_Vasnetsov_JModernOpt,1995_He_RubinszteinDunlop_JModOpt} and spatial light modulators \cite{2004_Gibson_FrankeArnold_OptExp,2004_Leach_Padgett_NewJPhys}. 
Similarly, polarization singularities typically require the additional integration of waveplates \cite{1990_Hajnal_PRSLA} or other birefringent media \cite{2010_Kurzynowski_Borwinska_JOpt,2010_Beckley_Alonso_OptExp}, although some geometries can also be realized with pinhole interferometers \cite{2009_Schoonover_Visser_PRA,2017_Ruchi_Senthilkumaran_OptExp}. Full polarization or spin singularities may further require the addition of tightly focusing optics such as lenses\cite{2022_Wang_Fan_Optica}. 

Metasurfaces, which comprise subwavelength-spaced nanostructures arrayed on a plane \cite{2000_Smith_Schultz_PRL,2001_Bomzon_Hasman_OptLett,2014_Yu_Capasso_NatMater,2022_Dorrah_Capasso_Science}, can reduce the complexity of the required system to a single surface. Their ability to manipulate and have optical functions that are contingent upon each degree of freedom of light (phase, incident wavevector, polarization state, spectrum) opens pathways to create optical singularities that were previously infeasible \cite{2019_Zhang_Gao_SciRep,2019_Ren_Genevet_NatComm,2021_Lim_Capasso_NatComms,2023_Lim_Capasso_NatComm,2023_Spaegele_Capasso_SciAdv}. In this review, we provide a framework for optical singularities in terms of their shape (geometry) and behavior (topology). Analogous singularity types can be categorized in a straightforward manner, potentially allowing the techniques developed for one type to find natural extension across other singularity classes. We show how metasurfaces and more generally wavefront-controlling surfaces can be incorporated into the generation and manipulation of these singular structures, and demonstrate how the unique mathematical properties of engineered optical singularities have enabled new applications in optics and beyond. 

\section{A brief history of optical singularities}

While new singular optical structures continue to emerge at the forefront of contemporary research, they have been a subject of study for many years. In 1919, V.S. Ignatowsky identified a strange phenomenon adjacent to a zero-intensity ring in the Airy disk of a focusing lens: in his diffraction calculation, energy appeared to be flowing backwards from the lens towards the source \cite{1919_Ignatowsky_TransOptInst,2019_Braat_Torok_ImagingOpt_Ignatowsky}. Richards and Wolf later verified these backflow predictions numerically \cite{1959_Richards_Wolf_PRSL}. Later, in 1950, H. Wolter, while studying the Goos-H\"{a}nchen beam shift in reflection, examined the fields surrounding a zero of the electric field and identified a circulation pattern in the phase profile around that point \cite{1950_Wolter_ZNaturforsch,2009_Wolter_JOA}. These zero field positions in diffraction calculations were first identified as field singularities with vortex behavior by Boivin, Dow, and Wolf in 1967 \cite{1967_Boivin_Wolf_JOSA}.

The identification of phase singularities as objects worthy of detailed study was initiated in 1974 by J.F. Nye and M.V. Berry \cite{1974_Nye_Berry_PRSA}. They provided a systematic mathematical description of phase singularities in a scalar field, which they identified as wave dislocations. These structures were serendipitously created five years later by J.M. Vaughan and D.V. Willetts, who discovered that certain modes of a laser operated at high power had a reduced on-axis intensity and an interference fringe profile that was well-explained by an ``Archimedean screw'' phasefront \cite{1979_Vaughan_Willetts_OptComm}. The term ``optical vortices'' was first used to describe these helical, circulating phase profiles around a phase singularity by P. Coullet in 1989 \cite{1989_Coullet_Rocca_OptComm}.

Scalar approximations to vectorial electromagnetic fields only hold when they are monochromatic, uniformly polarized, and paraxial (\ie propagating at small angles to the optic axis). Beyond phase singularities in this scalar wave approximation, electromagnetic fields with a nonuniform polarization profile also have singularities in their polarization parameters. In 1983, Nye proposed a polarization analogue to phase singularities by introducing wave disinclinations, a time-domain behavior in which the transverse polarization fields at a point in time $t$ vanish: $E_x(t)=0, E_y(t)=0$ \cite{1983_Nye_PRSLA_disinclinations}. Traced over time, these points sweep out a 2D surface, which he called the S surface. This 2D surface represents the set of linear transverse polarizations in a 3D time-harmonic field. The handedness of linear polarization on the S surface is undefined; S surfaces are thus singularities of polarization ellipse handedness. In the same year, Nye further identified loci comprising strictly circular transverse polarization, which he called C loci \cite{1983_Nye_PRSLA}. Generically, such ``C lines'' are 1D lines for fields with random intensity and phase in 3D space. Just as the globe longitudes and time zones intersect at the Earth's poles so that the longitudinal coordinate is undefined there, the polarization azimuth $\Psi$ is singular at the Poincar\'e sphere poles, at which right and left circular polarization reside. C loci are singularities of the polarization azimuth. Nye and Hajnal later generalized C loci and L loci (collections of points with undefined polarization ellipse handedness) to full vectorial fields (\ie fields with three scalar components representing the three Cartesian directions) to introduce $C^T$ lines, and $L^T$ lines where the $T$ represented ``True'' \cite{1987_Nye_Hajnal_PRSL}. Along $C^T$ lines, the vectorial polarization is strictly circular, and along $L^T$ lines the vectorial polarization is strictly linear. These vectorial polarization singularities were detected experimentally soon after by Hajnal in free-space microwave fields \cite{1990_Hajnal_PRSLA}.

\section{Generic vs. non-generic singularities}
We need to make a distinction here between generic \cite{dennis2001topological} and non-generic singularities. Generic singularities are stable; that is, they are robust to small perturbations in the field. These perturbations displace the singularity in its parameter space but do not destroy it. Thus these singularities persist in the physical world. The precise mathematical property of being generic is that the addition of a perturbation to the singular structure (in terms of the wavefront or wave generation conditions) produces a perturbed structure that is related to the original structure by a smooth transformation called a diffeomorphism\cite{1976_Berry_AdvPhys}. Conversely, non-generic singularities are either destroyed or split into multiple generic singularities by perturbations of arbitrarily small magnitude\cite{dennis2006rows}. These perturbations can come from stray light or device imperfections and to a good approximation, appear as additions or subtractions of plane waves in the neighborhood of the singularity. 

For instance, phase singularity points in 3D space \cite{2000_Arlt_Padgett_OptLett} are destroyed by the addition of stray light. Non-generic singularities require precise alignment of conditions that are almost never realized spontaneously in random fields. Random fields are unstructured superpositions of plane wave sinusoids that lack symmetries such a reflection axis or translational symmetry, among others. An example of a random field is the speckle pattern formed from diffuse laser light reflection off a rough surface.

Yet, both types of singularities have experimental relevance. Generic singularities are particularly useful in environments with high scattering, where their stability against perturbations or the existence of a perfect zero is crucial (\eg for high-precision sensing applications, details in Section \ref{sensing}). Non-generic singularities, on the other hand, are suitable in environments with low scattering and tight control over optical fields, or in situations where a deviation from a perfect mathematical zero is acceptable (\eg some optical communication applications, details in Section \ref{OAM}). Metasurfaces have produced both generic and non-generic singularities in experimental settings\cite{2019_Zhang_Gao_SciRep,2019_Ren_Genevet_NatComm,2021_Lim_Capasso_NatComms,2023_Lim_Capasso_NatComm,2023_Spaegele_Capasso_SciAdv}.

\section{Parameter space and condition space}

To fully design and exploit the topological nature of optical singularities, one has to fully understand their mathematical properties, geometrical structure, and insensitivity to imperfections (\ie topological protection). While the nomenclature can be initially confusing due to the range of singularity types, all optical singularities can be described by their \textit{parameter space} and \textit{condition space}. The parameter space of a singularity is the space that describes the \textit{location} of the singularity. For example, for the canonical phase singularity in the transverse plane, the parameter space is the two-dimensional space of $(x,y)$ spatial coordinates. While the Cartesian coordinates $(x,y,z)$ are the most common parameter space coordinates, singularities can also be defined in angular spaces (\eg for far-field projections) or in synthetic spaces like the incident wavelength $\lambda$ and tilt of illumination onto an optical device \cite{2018_Yuan_Fan_Optica,2023_Spaegele_Capasso_SciAdv}. For instance, spatiotemporal beams are those defined in a high-dimensional frequency-wavenumber space: $(\omega,\bm{k})$\cite{2012_Bliokh_Nori_PhysRevA}. The parameter dimension $D_{parameter}$, or domain dimension, is the dimension of this parameter space: the smallest number of unit vectors required to fully describe a position in the parameter space; it is 2 for the transverse $(x,y)$ plane and 3 for the full Cartesian $(x,y,z)$ space. 

The \textit{condition} space describes \textit{what is happening} at each position in parameter space and defines the existence of a singularity. Singularities are \textit{zeros} of condition space; the condition variables are simultaneously identically zero at the singularity. For example, the conditions for there to be a phase singularity are that the scalar field $E$ vanishes: $Re(E)=0, Im(E)=0$, so the condition space is $\{Re(E),Im(E)\}$, and the phase singularity exists at the $(0,0)$ origin of this condition space. The condition dimension $D_{condition}$, which is sometimes called the co-dimension (although ``co-dimension'' is not a contraction of ``condition dimension''), is the number of independent conditions (equations) that specifies the singularity. This can be qualitatively interpreted as the number of independent constraints imposed on the electromagnetic field or the number of independent variables needed to identify that structure\cite{2019_Berry_Shukla_JOpt}. While the parameter space dimension can be chosen by the designer, each class of electromagnetic wave singularities is connected to a predefined set of conditions (see Table \ref{table1}, discussed below), giving the condition space a fixed dimension. For instance, phase singularities have a fixed co-dimension of two, since the vanishing of the real and imaginary scalar field components are two independent conditions. A more precise mathematical formulation of optical fields as mappings between parameter and condition space is detailed in the Supplementary Information. 

We can describe the shape of a generic singularity by its dimension: whether it is a 0D point, 1D line, 2D surface or so on. This generic dimension is the shape that the singularity will take in a random wavefield and is given by the difference between the parameter dimension (domain dimension) and the condition dimension (co-dimension) in Equation \ref{eqn:singularitydimension}. This relationship is analogous to the dimension of a solution space in linear algebra; it is the number of algebraic unknowns (domain) minus the number of constraining equations (conditions). In linear algebra, if one has three unknowns and two constraint equations, the solution set lies along a $3-2=1$-dimensional line. 
\begin{gather}
D_{singularity} = D_{parameter} - D_{condition}\label{eqn:singularitydimension}
\end{gather}
More specifically, for a phase singularity in the transverse plane, the parameter dimension is 2 (for $x$ and $y$ coordinates) and the condition dimension is also 2 ($\{Re(E), Im(E)\}=\{0,0\}$ at the singularity). Therefore, the generic dimension is $2-2=0$, and phase singularities in the complex plane are 0D points. Note that if the parameter space is changed, then the generic dimension changes too. Consider phase singularities in 3D space (parameter dimension = 3). Generic phase singularities are thus $3-2=1$ dimensional lines in this space. This means that generic singularities do not have an inherent shape; \textit{their dimension is dependent on the parameter space they are defined in}. L type loci can be L points\cite{2002_Freund_OptComm}, L lines \cite{1987_Hajnal_PRSLA_I,1987_Hajnal_PRSLA_II,2002_Freund_OptComm}, and L surfaces \cite{2018_Shvedov_Krolikowski_NewJPhys}, depending on their embedding parameter space. 
 
Table \ref{table1} lists each optical singularity type, the mathematical conditions for their existence, and their corresponding co-dimension. Instead of using the variety of singularity names in literature, many of which presuppose a particular singularity dimension (\eg C-lines, V-points), we propose a simplified classification which focuses on the defining conditions of a singularity and the singular parameters. We still align closely to the earliest singularity parameter definitions introduced in the late 20th century, but explicitly make a distinction between transverse (paraxial) and full vectorial (non-paraxial) definitions. Paraxial fields have a well-defined direction of propagation and negligible longitudinal polarization, whereas the longitudinal fields are considered in non-paraxial fields. This distinction often gets lost in the naming conventions and depends on the context of the report. We do not include implied geometry labels, such as ``point'', ``line'', and ``surface'', since such a shape depends on the parameter dimension, as noted previously. There is further ambiguity in describing the dimension of a singularity: a 3D zero or 3D singularity could refer to either a point zero in a 3D space\cite{2023_Vernon_RodriguezFortuno_Optica}, a singularity with 3D structure\cite{2021_Droop_Denz_JOpt} but zero volume, or a volumetric (3D) space that has zero field value. These ambiguities necessitate a singularity naming system that is independent of the implied singularity geometry. 

\begin{table}[!t]
    \centering
    \caption[Proposed classification of singular features]{\label{table1} Classification of singular features, along with their defining mathematical properties and characteristic co-dimension (\ie number of conditions). $S_1=|E_x|^2 - |E_y|^2,S_2=2 \text{ Re}(E_xE_y^*)$ and $S_3=2\text{ Im}(E_xE_y^*)$ are three of the four Stokes parameters in the left-handed convention\cite{1993_Collett_PolarizedLight}, where $S_0=|E_x|^2+|E_y|^2$ is the paraxial field intensity.}
    \small
    \begin{tabular}{|p{2.2 cm}|p{5.2 cm}|p{5 cm}|c|c|}
    \hline
    \textbf{Type} & \textbf{Singular parameter} & \textbf{Conditions}  & \textbf{Co-dim.} & \textbf{Refs.} \\
    \hline
    Phase & $\arg(E)$ & $E\equiv Re(E)+i Im(E)=0$ &  2 & \cite{1974_Nye_Berry_PRSA} \\
    \hline 
    Disinclination & Instantaneous transverse polarization & $E_x(t)=0, E_y(t)=0$  & 2 & \cite{1983_Nye_PRSLA_disinclinations} \\
    \hline
    Transverse C & Transverse polarization azimuth or arg($S_1+iS_2$) & Transverse circularly polarized: $\bm{E}_\perp\cdot\bm{E}_\perp = E_x^2+E_y^2=0$, or equivalently, $S_1+iS_2 = 0$ & 2 &  \cite{1987_Hajnal_PRSLA_I,2013_Berry_JOpt} \\
    \hline
    Vectorial C & Vectorial polarization azimuth & Vectorial circularly polarized: $\bm{E}\cdot \bm{E} = E_x^2+E_y^2+E_z^2=0$  & 2 & \cite{1987_Nye_Hajnal_PRSL} \\
    \hline
   Transverse L & Transverse polarization ellipse handedness, which is sign($S_3$) & Transverse linearly polarized: $(\bm{E}_\perp,0)\times (\bm{E}^*_\perp,0)=0$, or equivalently, $S_3=0$  & 1 & \cite{1983_Nye_PRSLA_disinclinations,1987_Hajnal_PRSLA_I,1987_Hajnal_PRSLA_II,2002_Freund_OptComm} \\
    \hline
    Vectorial L & Azimuth of polarization ellipse normal vector & Vectorial linearly polarized: $\bm{E}\times \bm{E}^* =0$  & 2 & \cite{1983_Nye_PRSLA_disinclinations,1987_Nye_Hajnal_PRSL} \\
    \hline
    Transverse V & Transverse polarization ellipse & $\bm{E}_\perp\equiv (E_x,E_y)=\bm{0}$  & 4 & \cite{2001_Freund_OptComm,2002_Freund_OptComm} \\
    \hline
    Vectorial V & Vectorial polarization ellipse & $\bm{E}\equiv (E_x,E_y,E_z)=\bm{0}$  & 6 & \cite{2000_Arlt_Padgett_OptLett,2006_Bokor_Davidson_OptLett,2006_Kozawa_Sato_OptLett} \\
    \hline
    Spin defect & Spin angular momentum density & $\bm{S}=(1/4\omega)[\epsilon_0 Im(\bm{E}^*\times \bm{E})+\mu_0 Im(\bm{H}^*\times \bm{H})]$$\equiv(\sigma_x,\sigma_y,\sigma_z)=\bm{0}$  & 3 & \cite{2022_Wang_Fan_Optica} \\
    \hline
    Cross-spectral density & $\arg(\mu)$ & $\mu(r_1,r_2;\omega)=\langle E^*(r_1, \omega) E(r_2, \omega)\rangle=0$  & 2 & \cite{2003_Schouten_Wolf_OptLett} \\
    \hline
    Mutual coherence function & $\arg(\Gamma)$ & $\Gamma(r_1,r_2;\tau)=\langle E^*(r_1, t) E(r_2, t+\tau)\rangle =0$  & 2 & \cite{2004_Maleev_Swartzlander_JOSAB} \\
    \hline
    \end{tabular}
    
\end{table}

\begin{figure}[!t]
\centering
\includegraphics[width=15 cm]{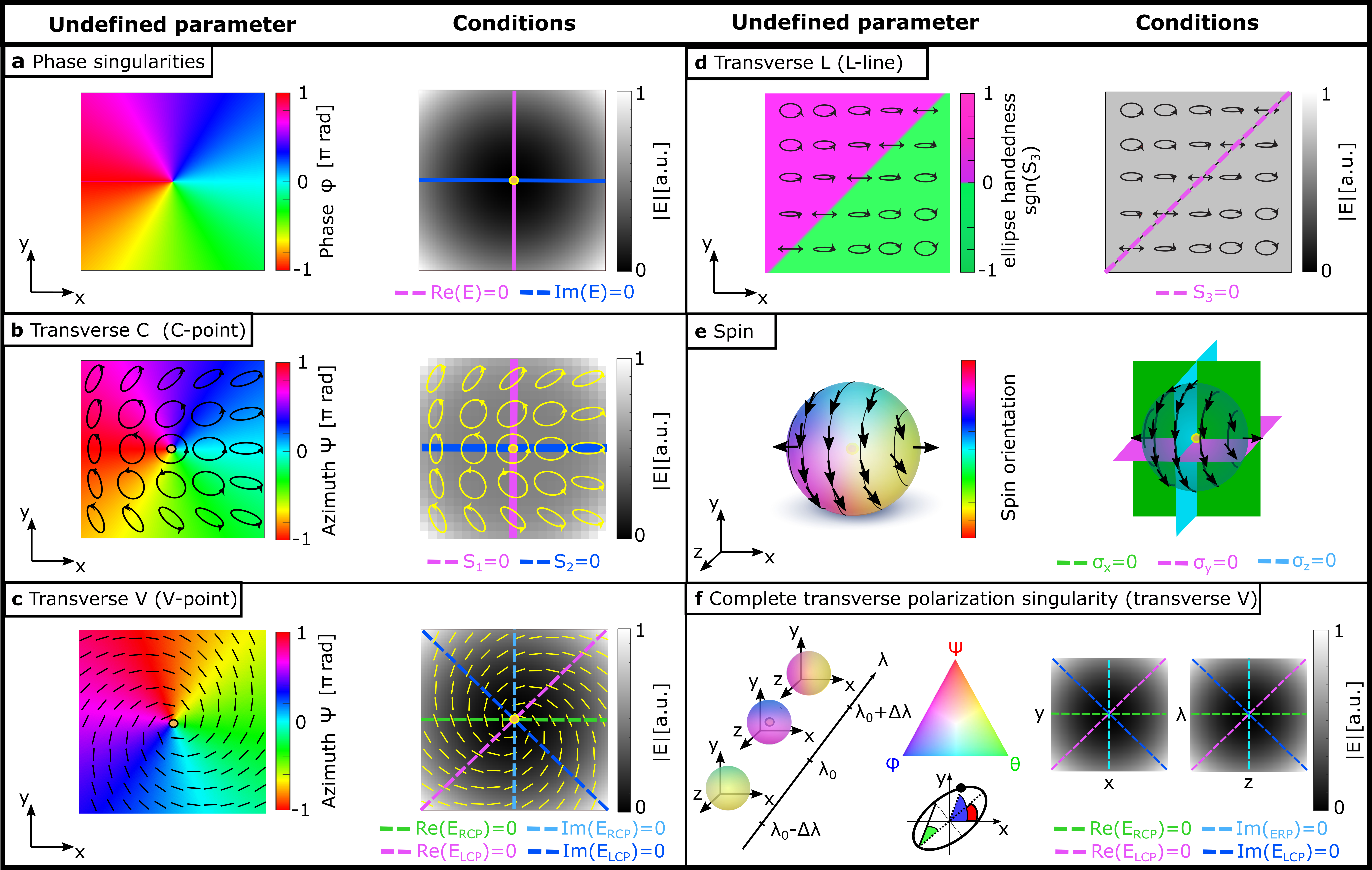}
\caption[]{Representative geometries of common singularity types, detailing the associated singular parameters and the conditions required for their existence. Singularities occur at the origin of condition space, so they exist where all the condition values are simultaneously zero. a.u. indicates arbitrary units. \textbf{a} Phase singularities occur in a scalar field $E$ when the amplitude vanishes $|E|=0$, which has a co-dimension of two since both the real and imaginary parts of $E$ vanish. This produces an intensity zero and generically a vortex-like phase swirl in the transverse plane. \textbf{b} Transverse C loci (commonly known as C-points) occur where the polarization azimuth (angle of polarization ellipse major axis) is undefined and hence are points of pure circular polarization. They have a co-dimension of two since the $S_1=|E_x|^2 - |E_y|^2$ and $S_2=2 \text{ Re}(E_xE_y^*)$ components of the Stokes vector have to vanish. \textbf{c} Transverse V loci occur where the transverse complex electric field $E_x,E_y$ is zero, hence have a co-dimension of 4. \textbf{d} Transverse L loci occur where the transverse polarization ellipse handedness (clockwise or counterclockwise orientation of the polarization ellipse) is undefined and are points of linear polarization. These form lines in the transverse plane due to their co-dimension of 1. The ellipse handedness is the sign of the $S_3=2\text{ Im}(E_xE_y^*)$ Stokes parameter. \textbf{e} Spin defect singularities occur where the spin angular momentum density $\bm{\sigma}=(\sigma_x,\sigma_y,\sigma_z)$ is zero \cite{2022_Wang_Fan_Optica}. Picture adapted with permission from H. Wang, C.C. Wojcik, and S. Fan\cite{2022_Wang_Fan_Optica}. \copyright Optica Publishing Group. \textbf{f} A topologically protected polarization singularity \cite{2023_Spaegele_Capasso_SciAdv} is similar to a V-point, but is defined in a higher-dimensional 4D space so that it exists as a stable 0D singularity. The immediate neighborhood of the singularity contains exactly one instance of each polarization state (parametrized by the azimuth $\Psi$ and ellipticity angle $\theta$), including the global phase $\phi$, which is proportional to the sum of the individual electric field phases $\phi_x+\phi_y$.
\label{fig1}}
\end{figure}
Figure \ref{fig1} plots representative geometries of the most common singularity types, their undefined (singular) parameters, and the conditions that create each singularity. In each subfigure, we plot the field distribution in the vicinity of the singularity on the left and show where the individual conditions are satisfied on the right. Singularities occur when all the condition values are simultaneously zero, so they occur at the intersection of the zero-contour lines or zero-isosurfaces of these individual conditions.

Phase singularities (Figure \ref{fig1}a) are the most well-known singularities, and occur when the real and imaginary parts of a scalar field simultaneously vanish, producing positions of undefined phase. The condition values for phase singularities are thus $\{Re(E),Im(E)\}$. 

C singularities (Figure \ref{fig1}b) are singularities in the polarization state of light at which the polarization azimuth is singular. The polarization azimuth is the orientation of the major axis of the polarization ellipse, and it is singular for circularly polarized light. The field at the center of a C singularity is thus strictly circular, hence its name. C singularities are either transverse (paraxial, no longitudinal field) or fully vectorial (longitudinal field included). Transverse C singularities are zeros of the complex Stokes field $\Sigma = S_1+iS_2$ \cite{2002_Freund_OptComm}, where $S_1$ and $S_2$ are the first two Stokes parameters (see Supplementary Information for the mathematical form). The condition values for transverse C singularities are thus $\{S_1,S_2\}$. 

V singularities are positions at which the full polarization state is undefined; they are positions of zero vectorial electric field. Transverse V singularities have zero transverse fields $E_{x,y}=0$, and vectorial V singularities have zero longitudinal field as well $E_{x,y,z}=0$. Since each of the time-harmonic fields can be represented by a complex number, the condition values for transverse V singularities are $\{Re(E_x),Im(E_x),Re(E_y),Im(E_y)\}$ and that of vectorial V singularities are $\{Re(E_x),Im(E_x),Re(E_y),Im(E_y),Re(E_z),Im(E_z)\}$. A proper subset of transverse V singularities is V-points \cite{2002_Freund_OptComm} (Figure \ref{fig1}c), which are positions of zero transverse field surrounded by strictly linear polarization and originally defined in the transverse $(x, y)$ plane. However, as we will discuss in a later section, those singularities are not stable due to the mismatch of parameter- and condition space dimensions, splitting the V point into multiple stable C loci upon perturbation \cite{2021_Liu_Kivshar_Nanophoton}. Recently, Spaegele et al created a stable transverse V singularity defined in a 4-dimensional parameter space \cite{2023_Spaegele_Capasso_SciAdv} (Figure \ref{fig1}f). The parameter space for this singularity included the three Cartesian directions and the wavelength of incident light. This singularity, which was also known as a complete transverse polarization singularity, realized every transverse polarization state, including the global phase, in the 4-dimensional vicinity of the singular point. 

L singularities are positions at which the handedness of the polarization ellipse is undefined. Transverse L singularities (Figure \ref{fig1}d) are commonly known as L-lines in the transverse plane\cite{1987_Nye_Hajnal_PRSL,2002_Freund_OptComm}. For the transverse polarization ellipse, the handedness is determined by sign of the $z$-directed spin angular momentum, which is proportional to the $S_3$ Stokes parameter. $S_3$ serves as the only condition value for this $D_{condition}=1$ system, making it a line in the transverse $(x,y)$ plane. The sign of $S_3$ is undefined at $S_3=0$, at which the singularity resides.

H. Wang, C.C. Wojcik and S. Fan recently introduced a spin defect singularity\cite{2022_Wang_Fan_Optica}  with $D_{condition}=3$ (Figure \ref{fig1}e). 
This is the position in space where the total spin angular momentum (SAM) density of light, $\bm{\sigma} = (1/4\omega)[\epsilon_0 Im(\bm{E}^*\times \bm{E})+ \mu_0 Im(\bm{H}^*\times \bm{H})]$, vanishes. Since the total spin angular momentum density is a vector with three independent elements, this singularity corresponds to $D_{condition}=3$, where the condition values are each of the SAM density components $\{\sigma_x,\sigma_y,\sigma_z\}$. 

\begin{figure}[!t]
\centering
\includegraphics[width=8
cm]{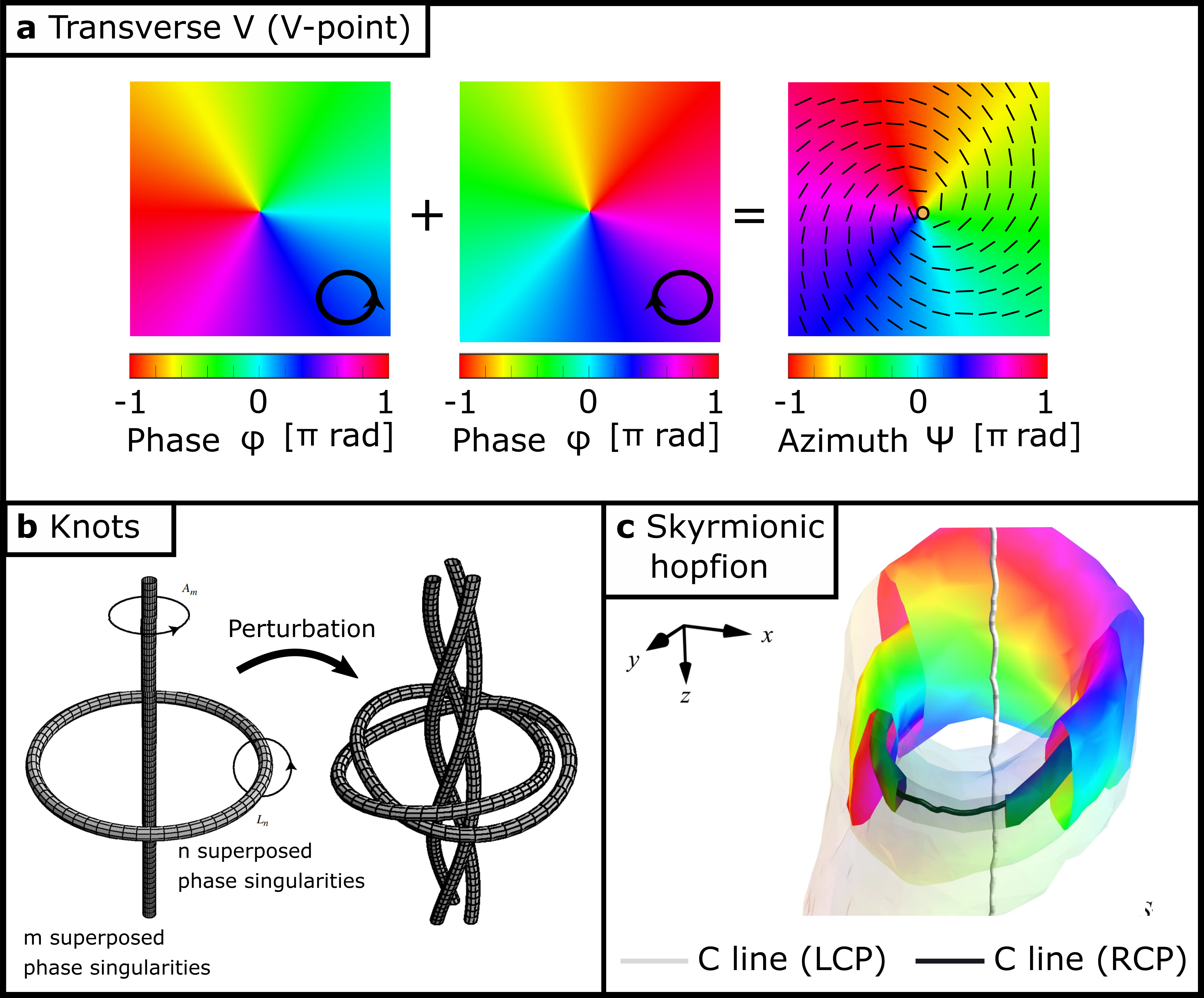}
\caption[]{Examples of singular structures formed at or around compositions of fundamental
singularities. \textbf{a} V-points can be understood as a superposition of phase singularities in left-hand circular (LCP) and right-hand circular (RCP) polarized light.  \textbf{b} Example of a torus knot threaded by a helix, that can be created through the perturbation of multiple superposed phase singularity lines and rings. Used with permission of The Royal Society (U.K.), from ``Knotted and linked phase singularities in monochromatic waves'', M.V. Berry and M.R. Dennis, Proc. Royal. Soc. A 457, 2013  (2001)\cite{2001_Berry_Dennis_ProcRoyalSocLonA}; permission conveyed through Copyright Clearance Center, Inc. \textbf{c} Skyrmionic hopfions can be understood as a C-line threading another closed C-loop of opposite handedness. Image from D. Sugic et al \cite{2021_Sugic_Dennis_NatComm} under the Creative Commons CC-BY license.}
\label{fig2}
\end{figure}
Beyond phase and polarization singularities, coherence singularities also exist, albeit having garnered significantly less attention. These are singularities in the degree of correlation of electromagnetic fields, which is also referred to as coherence \cite{1938_Zernike_Physica,1965_Mandel_Wolf_RevModPhys}. In 2003, Schouten et al identified singularities in the Cross-Spectral density \cite{2003_Schouten_Wolf_OptLett}, and in 2004, Maleev et al characterized singularities in the Mutual Coherence Function \cite{2004_Maleev_Swartzlander_JOSAB}. Both are singularities in the complex phase of a scalar field, and thus have co-dimension of 2. 

Multiple other structures in singular optics are formed at or around compositions or variations of these fundamental singularities displaced in parameter space. They include, among others, Stars, Monstars, Lemons, and V-points \cite{1983_Nye_PRSLA} (superposed C-points, Figure \ref{fig2}a), knots \cite{2001_Berry_Dennis_ProcRoyalSocLonA,2010_Dennis_Padgett_NatPhys} (superposed phase-lines with a plane wave perturbation, Figure \ref{fig2}b), 3D Skyrmions \cite{2023_Ehrmanntraut_Denz_Optica,2021_Shen_Zheludev_NatComm,2018_Tsesses_Bartal_Science,2022_Parmee_Ruostekoski_CommsPhys} (phase singularity line(s) in $E_x$ threading phase singularity loop(s) in $E_y$), skyrmionic hopfions \cite{2021_Sugic_Dennis_NatComm} (one realization involves a C-line threading another closed C-loop of opposite handedness, Figure \ref{fig2}c), and optical M\"{o}bius strips \cite{2015_Bauer_Leuchs_Science,2019_Tekce_Denz_OptExp} (half-integer twisting of a polarization major axis around tightly-focused polarization singularities). The space of composite singularities is rich and complex, but we will focus on the fundamental constituent singularities here. We commend the review of Shen et al\cite{2023_Shen_Zayats_NatPhoton} on optical skyrmions to the interested reader.

Table \ref{table2} shows an overview of each singularity class mentioned, organized by the dimensions of parameter and condition space. Different cell colors correspond to different generic dimensions: Orange indicates points (0D), blue indicates lines (1D), yellow indicates surfaces (2D), and green indicates 3D or higher dimensional geometries. 

Most works have been restricted to the parameter space of the Cartesian spatial dimensions $(x,y,z)$, which cannot create stable singularity structures with co-dimensions larger than 3, such as stable transverse polarization singularities and stable non-paraxial polarization singularities. Singularities with a condition dimension larger than the parameter dimension have no stable geometry (Table \ref{table2} cells with a white background). This means that although such singularities can be designed by precise alignment of system symmetries, they are inherently vulnerable to destruction by arbitrarily small perturbations, precluding their experimental realization in a strict mathematical sense \cite{2002_Freund_OptComm,2006_Schoonover_Visser_OptExp,2016_Otte_Denz_JOpt}. Singularities can only be protected against perturbations if the dimension of the parameter space is larger or equal to that of the condition space\cite{2023_Spaegele_Capasso_SciAdv}. This gives rise to the first design rule when sculpting singularities: To design a generic singularity of dimension $d$ that is stable against perturbations, one must choose $D_{parameters}=D_{conditions}+d$. Synthetic dimensions (parameters of the system such as angle of incidence or the frequency of the light \cite{2018_Yuan_Fan_Optica}) are viable ways to increase the parameter space dimensions to achieve high-dimensional singularities. The transverse V singularity (co-dimension of 4) is stable if it is defined in a higher dimensional 4D space, such as that of 3D spatial dimensions and one synthetic dimension\cite{2018_Yuan_Fan_Optica} of wavelength, where it exists as a stable generic point (0D) singularity\cite{2023_Spaegele_Capasso_SciAdv}. In the case where the sources of perturbation can be mitigated or tightly controlled (\ie high device fabrication precision, no stray light or scattering media), one can also choose to design singularities that are not protected through their topology. While the field will not achieve perfect zeros for each individual condition value in experimental reality, they can still be engineered to be sufficient (in terms of intensity contrast, surrounding field distribution, or confinement) for the desired applications. 

\begin{table}[ht]
    \centering
    \caption[Relationship between singularity dimension, parameter space dimension and number of conditions]{The singularity dimension is determined by the difference of parameter space dimension and the number of conditions that need to be fulfilled. The background cell colors indicate the respective dimension of generic singularities with that combination of parameter dimension and singularity co-dimension. Orange indicates points (0D), blue indicates lines (1D), yellow indicates surfaces (2D), and green indicates 3D or higher dimensional geometries. Cells with a white background do not have topologically protected singularities as the condition space dimension exceeds the parameter space dimension. Singularities with parameter space dimension exceeding 3 can use synthetic dimensions due to there only being three Cartesian dimensions.}
    \begin{tabular}{|p{1.5cm}|p{2cm}|p{2.5cm}|p{2cm}|p{2cm}|p{2cm}|p{2.5cm}|}
    \hline
     & Co-dim=1 & Co-dim=2 & Co-dim=3 & Co-dim=4 & Co-dim=5 & Co-dim=6 \\
    \hline
    Parameter dim=1 & \cellcolor[rgb]{0.9,0.62,0} & & & & & \\
    \hline 
    Parameter dim=2 & \cellcolor[rgb]{0.34,0.71,0.91} L line\cite{2002_Freund_OptComm}& \cellcolor[rgb]{0.9,0.62,0} Phase vortex\cite{1974_Nye_Berry_PRSA}, c-point\cite{1987_Hajnal_PRSLA_II,2002_Freund_OptComm,2013_Berry_JOpt}, d point\cite{1987_Hajnal_PRSLA_II} & & & & \\
    \hline 
    Parameter dim=3 & \cellcolor[rgb]{0.94,0.89,0.26} S-surface\cite{1987_Hajnal_PRSLA_I} &\cellcolor[rgb]{0.34,0.71,0.91} C-line\cite{1987_Hajnal_PRSLA_I}, $C^T$ line\cite{1987_Nye_Hajnal_PRSL}, $L^T$ line\cite{1987_Nye_Hajnal_PRSL} & \cellcolor[rgb]{0.9,0.62,0} Spin defect\cite{2022_Wang_Fan_Optica}& V-point\cite{2001_Freund_OptComm,2002_Freund_OptComm}& & 3D hollow spots\cite{2006_Bokor_Davidson_OptLett,2006_Kozawa_Sato_OptLett,2007_Bokor_Davidson_OptComm,2011_Xue_Liu_JOpt,2012_Li_Liu_JOpt,2014_Wan_Qiu_LaserPhotRev,2018_Wu_Chen_OptExp,2023_Lim_Capasso_NatComm}\\

    \hline 
    Parameter dim=4 & \cellcolor[rgb]{0,0.62,0.45} & \cellcolor[rgb]{0.94,0.89,0.26} &\cellcolor[rgb]{0.34,0.71,0.91} & \cellcolor[rgb]{0.9,0.62,0} Phase/Pol. singularity\cite{2023_Spaegele_Capasso_SciAdv} & & \\
    \hline 
    Parameter dim=5 & \cellcolor[rgb]{0,0.62,0.45} & \cellcolor[rgb]{0,0.62,0.45} & \cellcolor[rgb]{0.94,0.89,0.26} &\cellcolor[rgb]{0.34,0.71,0.91} & \cellcolor[rgb]{0.9,0.62,0} & \\
    \hline 
    Parameter dim=6 & \cellcolor[rgb]{0,0.62,0.45}& \cellcolor[rgb]{0,0.62,0.45} & \cellcolor[rgb]{0,0.62,0.45} & \cellcolor[rgb]{0.94,0.89,0.26} &\cellcolor[rgb]{0.34,0.71,0.91} & \cellcolor[rgb]{0.9,0.62,0}\\
    \hline
    \end{tabular}
    \begin{minipage}{\textwidth}
      \includegraphics[width=\linewidth]{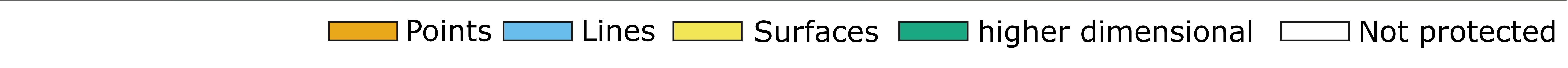}
    \end{minipage}
    \label{table2}
\end{table}
\section{Singularity stability and charges}

Given the beneficial robustness and stability of generic singularities, we seek a simple criterion to differentiate between generic and non-generic singularities. Beyond the requirement that the generic singularity dimension be equal to the difference between the parameter dimension and condition dimension, generic singularities must have a topological charge of $\pm 1$ \cite{1993_Basistiy_Vasnetsov_OptComm}. This section connects topological charge to the properties of the field around the singularity and singularity stability. 

\begin{figure}
\centering
\includegraphics[width=17 cm]{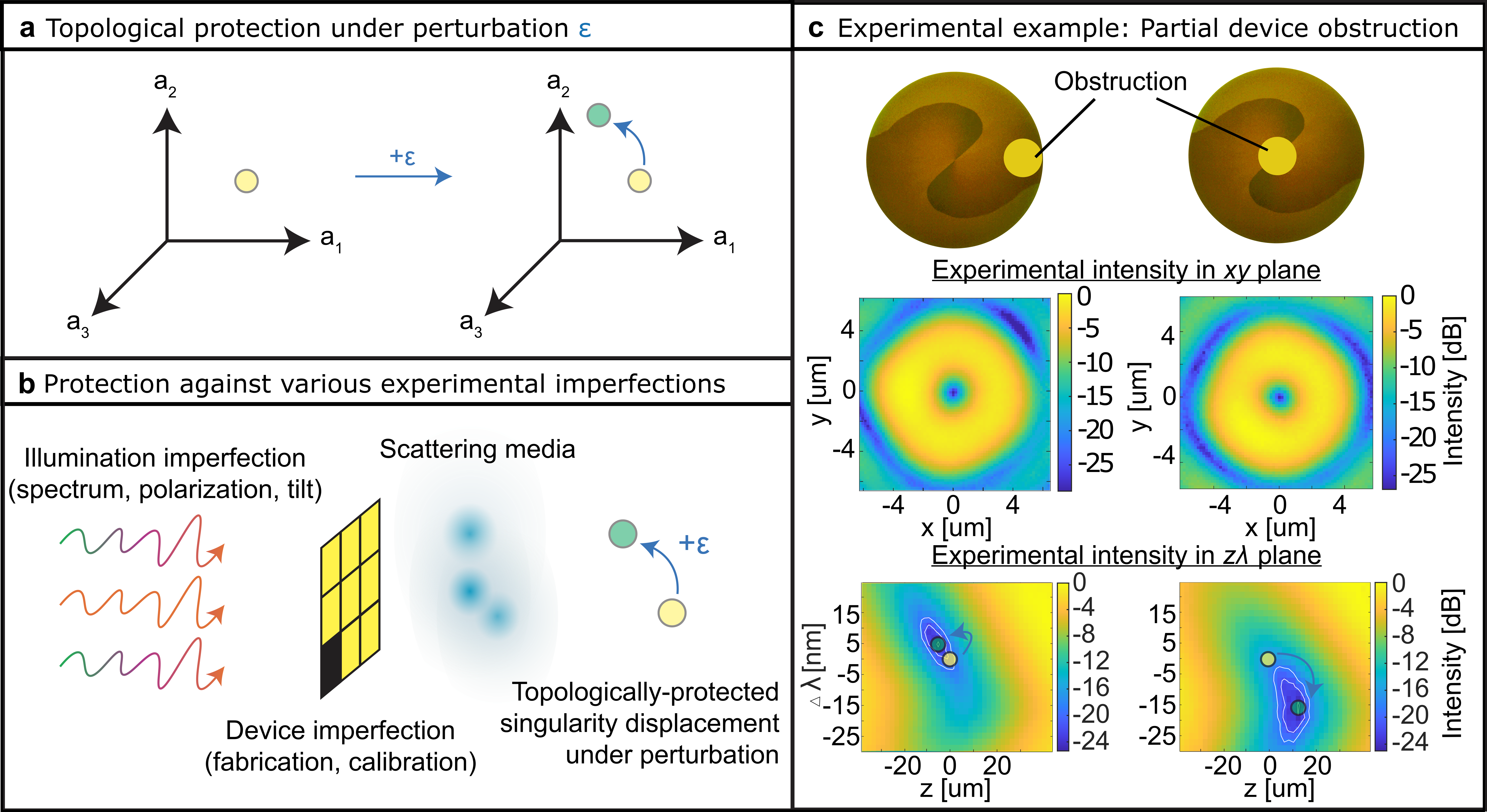}
\caption[]{
\label{fig3} \textbf{a} Topological protection refers to the persistent existence of a singularity under a small perturbation $\epsilon$ of the conditions, depicted here as $\{a_1,a_2,a_3\}$. $\epsilon$ is an additive shift in the condition values, such as stray light, which is an additive shift in the complex field. Topologically protected singularities are merely displaced in space under such a perturbation and not destroyed. \textbf{b} Perturbations can arise from various sources, including experimental imperfections. Imperfections in the state of illumination for an optical device, in terms of its spectral content, polarization state, or geometrical tilt perturbs a system from its desired state, introducing or subtracting fields at the singular positions. Imperfections in device design and fabrication also have the same perturbing effect, as does weakly scattering or absorbing propagation media. \textbf{c} Experimental demonstration of device imperfection on the perturbation of a topologically-protected polarization singularity defined in $(x,y,z,\lambda)$ parameter space from Spaegele et al\cite{2023_Spaegele_Capasso_SciAdv}. Part of a metasurface that generated the polarization singularity was obstructed by an opaque disk in two locations over two experiments. The singularity continued to exist under perturbation and was observed to be displaced primarily in $z\lambda$ space from its design position. From C.M. Spaegele et al, Topologically protected optical polarization singularities in four-dimensional space. Sci. Adv. 9, eadh0369 (2023)\cite{2023_Spaegele_Capasso_SciAdv}. \copyright The Authors, some rights reserved; exclusive licensee AAAS. Distributed under a Creative Commons Attribution NonCommercial License 4.0 (CC BY-NC) \href{http://creativecommons.org/licenses/by-nc/4.0/}{http://creativecommons.org/licenses/by-nc/4.0/}}
\end{figure}

Singularity stability, or topological protection, is the continued existence of the singularity under the influence of a small perturbation $\epsilon$ in the fields or condition values. This results in topologically protected singularities being displaced in parameter space but not destroyed (Figure \ref{fig3}a). Critically, such behavior occurs because a generic singularity has all possible condition values (which are usually field values), up to normalization, in the vicinity of the zero. These values can cancel out arbitrary (but small) condition perturbations. They must only occur once in the vicinity of the zero --- higher order singularities of charge $m$ exhibit each set of condition values appearing $m$ times in its vicinity, causing a splitting into $m$ generic singularities upon perturbation \cite{dennis2006rows}.

Singularity stability holds up to a limit. Each topologically-protected singularity is surrounded by all condition values up to a normalization scale, which means that a perturbation with magnitude larger than the surrounding field values can cause the singularity to disappear. The ``strength'' of the topological protection is determined by the maximum perturbation magnitude for which the singularity can enforce destructive interference. For example, a phase singularity is surrounded by all $Re(E)$ and $Im(E)$ values normalized by the scalar field magnitude $\sqrt{Re(E)^2+Im(E)^2}$. In other words, the phase singularity is surrounded by all values of $Re(E)/\sqrt{Re(E)^2+Im(E)^2}$ and $Im(E)/\sqrt{Re(E)^2+Im(E)^2}$, which is $\cos\phi$ and $\sin\phi$ for $arg(E)=\phi$, respectively. The protection ``strength'' of the phase singularity is determined by the largest field magnitude range $\sqrt{Re(E)^2+Im(E)^2}$ for which all phases exist in the vicinity of the singularity. A perturbation with a field magnitude that exceeds the maximum in that range will destroy that singularity. It is hence desirable for topologically protected singularities to be protected over a wide range of perturbation magnitude scales.

Beyond direct perturbations to optical singularities in terms of stray light contributions, perturbations to singular systems can also arise from indirect sources, such as imperfections in the illumination conditions, device design, and propagation media (Figure \ref{fig3}b). These deviations from ideal conditions have the effect of adding and subtracting fields at singular positions. Topologically-protected singularities are thus also protected against small imperfections in these conditions, exhibiting displacements in parameter space in response to these changes instead of being destroyed. This robustness against device imperfections was demonstrated experimentally in a study by Spaegele et al, in which a topologically-protected polarization singularity defined in a 4D $(x,y,z,\lambda)$ parameter space was perturbed by obscuring part of the metasurface that generated the singularity (Figure \ref{fig3}c). The singularity persisted upon obstruction with a displacement primarily in $z\lambda$ space\cite{2023_Spaegele_Capasso_SciAdv}.

Two primary methods can be employed to determine if the topological charge is $\pm 1$ and hence if the singularity is topologically protected. The first is the degree of the function, which is more commonly known as the winding number in 2D, and the second method uses the field gradient, specifically, the sign of the Jacobian determinant. 

\begin{figure}
\centering
\includegraphics[width=15 cm]{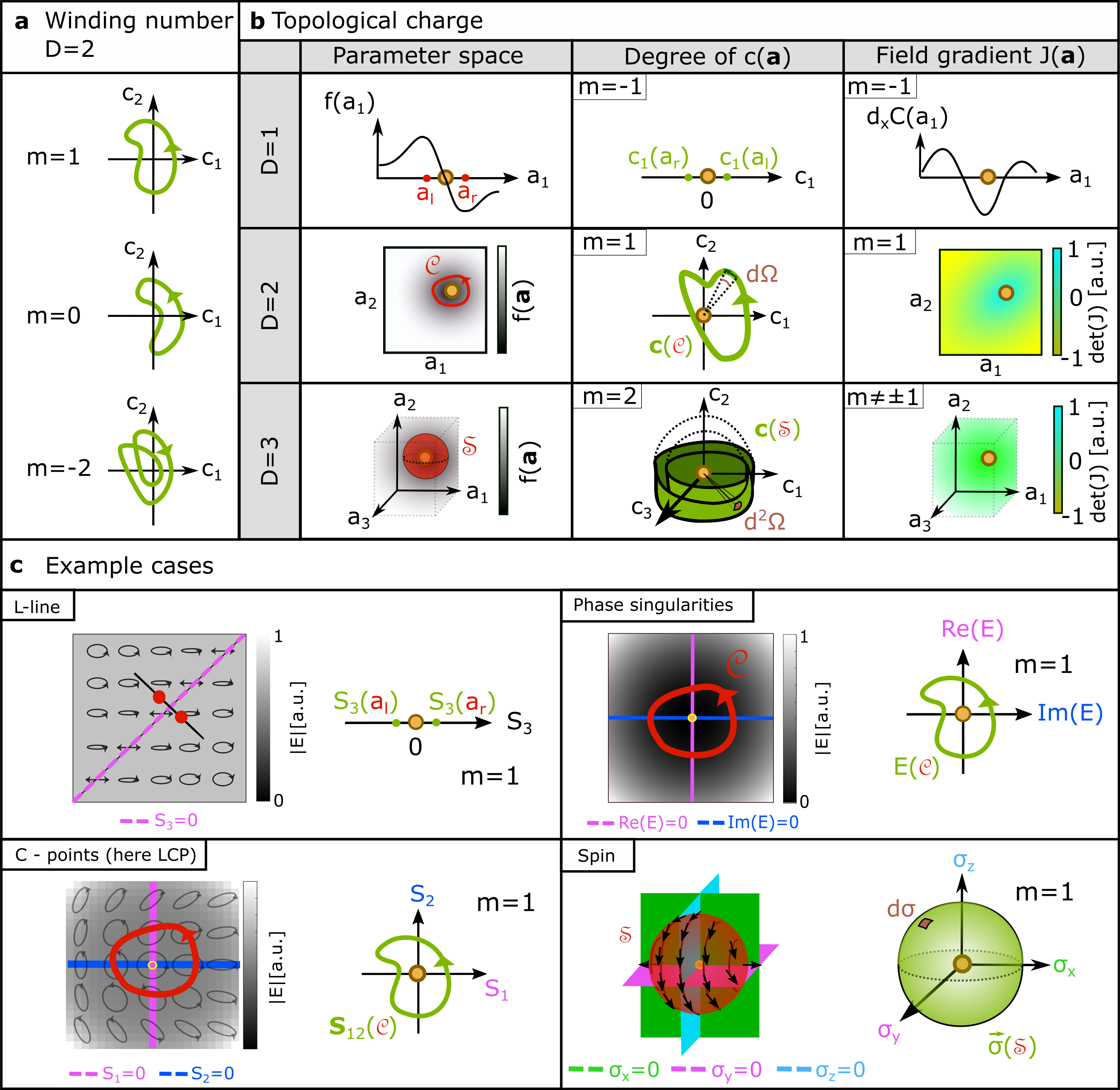}
\caption[]{
\label{fig4} Generalized topological protection and topological charge for singularities defined in arbitrary dimensions. Here, we only consider 0D point singularities with $D_{parameter}=D_{condition}=D$. \textbf{a} The winding number summarizes the behavior of an optical field around a singularity in the plane $D=2$. When encircling a point singularity on a plane, the corresponding trajectory in $\{c_1,c_2\}$ condition space (green loop) will encircle the condition space origin an integer number of times, which is the winding number. The winding number is the topological charge of the singularity on a plane. \textbf{b} For other dimensions, the topological charge is computed analogously to the winding number using the degree of a continuous mapping. This panel shows how the singularity is located in parameter space (first column), how the topological charge (degree of mapping) is computed (second column), and how the field gradients (parametrized by the magnitude of the mapping Jacobian determinant det($J$)) are extremal in the vicinity of the singularity (third column). ``a.u.'' indicates arbitrary units. For $D=1$, the singularity is ``encircled'' by two points $a_l$ and $a_r$ on its left and right. The condition values at these two positions $c_1(a_l),c_1(a_r)$ determine the charge of the singularity. For $D=2$, the topological charge corresponds to the conventional winding number. The singularity in $\{a_1,a_2\}$ space is encircled by a red loop $C$, which corresponds to the green condition space trajectory. For $D=3$, the singularity is ``encircled'' by a spherical shell $S$, which corresponds to another shell in condition space. The number of times the condition space shell ``wraps'' around the origin is the topological charge. \textbf{c} Exemplar topological charge calculations for common optical singularities. Transverse L loci (``L-lines") have $D_{condition}=1$ with a characteristic charge that corresponds to the jump in the sign of the $S_3$ Stokes parameter across the singularity, at which $S_3=0$. The topological charges of phase singularities and transverse C loci (both with $D_{condition}=2$) are computed using the winding number on the condition value plane. The spin defect singularity \cite{2022_Wang_Fan_Optica} has $D_{condition}=3$ and hence is computed by the degree of a continuous mapping between a thin spherical shell in parameter space to the corresponding thin shell in $\vec{\sigma}\equiv\{\sigma_x,\sigma_y,\sigma_z\}$ spin space.}
\end{figure}
\subsection{Ensuring stability using the degree of the function}
While there are several equivalent names for the topological charge in various contexts (\eg vortex charge, topological invariant, rotation number, Hopf index), they perform the same mathematical calculation (up to scaling) known in topology as the degree of the mapping between parameter and condition space\cite{2007_Schwarz_Topology}.

The degree of a function is most easily recognized in a 2D setting, where it is known as the winding number\cite{2009_Dennis_Padgett_ProgOpt}. This winding number is an integer that counts the number of times a curve loops around a point, such as the origin of condition space. The count can be either positive or negative, which is determined by the orientation of the curve --- \ie whether it winds the point in a counterclockwise (positive) or clockwise (negative) manner. Figure \ref{fig4}a presents various examples of curves with their corresponding winding numbers. 

The winding number of an optical singularity in 2D is computed similarly as follows: encircle the singularity in parameter space with a counterclockwise curve and track the corresponding trajectory within condition space, as depicted in the middle row of Figure \ref{fig4}b. The winding number of the closed oriented loop in condition space is the integer-valued degree of the singularity function. 

For phase singularities in 2D, the winding number $m$ can also be calculated by the well-known integral in Equation \ref{eqn:phase_charge} which accumulates the complex phase over a counterclockwise loop $C$ around the singularity in parameter space: 
\begin{gather}
m = \frac{1}{2\pi}\oint_C d\phi = \frac{1}{2\pi}\oint_C \nabla \phi \cdot d\bm{r}, m\in\mathbb{Z} \label{eqn:phase_charge}
\end{gather}

This integral is functionally identical to tracking the winding number of the trajectory in $\{Re(E),Im(E)\}$ condition space. The value of $m$ must be an integer, as a closed path returning to the starting position would yield a phase accumulation that does not equal $0$ modulo $2\pi$ otherwise. The orientation of the closed path determines the sign of the topological charge; retracing the same path in the opposite orientation will yield the same phase accumulation with the reverse sign. This quantity remains unchanged when the path is continuously translated or distorted in parameter space, and only alters its value by integer increments when phase singularities move into or out of the interior of the path.

For the canonical phase singularity in the transverse $xy$ plane, the topological charge of a singularity located at $(x_0,y_0)$ quantifies the arrangement of condition space values $\{Re(E(x,y)),Im(E(x,y)\}$ around that point (upper right in (Figure \ref{fig4}c)). The mapped curve wraps once around the origin of condition space, corresponding to $m=1$ with all condition values (up to normalization) appearing once in the immediate vicinity of the singularity.

This mechanism for topological protection generalizes to arbitrary dimensions of parameter and condition space\cite{2007_Schwarz_Topology}: as long as the map from the immediate vicinity of the singularity in parameter space achieves all possible values in condition space once (up to normalization), we say that the singularity is topologically protected. For simplicity, we will specialize here to point (0D) singularities, which only generically occur when $D=D_{parameter}=D_{condition}$; this restriction allows us to identify correspondences with common concepts in topology.

With 1D condition space, singularities are the zeros of a 1D function $f(a_1)$ (Figure \ref{fig4}b, first row). The topological charge of the zero is the change in function sign across the zero, or equivalently, the sign of the function slope at the zero. With a 2D condition space $\{c_1,c_2\}$ (as in the phase singularity example) we encircle the singularity in parameter space with a counterclockwise curve $C$ and count how often the corresponding trajectory in condition space wraps around the origin (Figure \ref{fig4}b, middle row). With 3D parameter spaces, we have to consider a spherical shell $S$ around the singularity in parameter space and calculate the number of wrappings that the condition values on this surface make around the origin of the condition space (Figure \ref{fig4}b, bottom row). Mapping degrees in higher dimensions are difficult to visualize, as they measure how often a ($D_{condition}-1$)-dimensional surface wraps around the origin of the $D_{condition}$-dimensional condition space\cite{2007_Schwarz_Topology}, but they can be calculated using Equation \ref{eqn:generalcharge}.
\begin{gather}
    l = \frac{1}{A}\oint_S d^{D_{condition}-1}\Omega, l\in\mathbb{Z} \label{eqn:generalcharge}
\end{gather}

$d^{D_{condition}-1}\Omega$ is a differential solid angle in $D_{parameter}=D_{condition}$ dimensions (evaluated over condition space values) and $A$ is the total solid angle of a $(D_{condition}-1)$-sphere (\ie the surface of a $D_{condition}-1$-dimensional sphere in parameter space). Notice that this reduces to the familiar closed loop topological charge for $D_{parameter}=D_{condition}=2$, since a 1-sphere is a circle with planar angles, $d^{2-1}\Omega = d\phi$, which has a total angle of $A=2\pi$ radians. 

Figure \ref{fig4}c exhibits the mechanism of determining the topological charge for several representative singularities. The topological charge is always determined by enclosing the singularity with a $(D_{condition}-1)$-sphere. For $D_{condition}=1$, we consider transverse L singularities (Figure \ref{fig4}c, top left), for which the topological charge is the change in $S_3$ sign across the singularity $l_L = S_3(a_r)-S_3(a_l)$, where $a_r$ and $a_l$ are points bounding the L locus. For 2D, the topological charges for both phase singularities (Figure \ref{fig4}c, top right) and transverse C singularities in the transverse plane (Figure \ref{fig4}c, bottom left) are computed in an identical manner through the winding number of a loop described earlier. The $D_{condition}=2$ condition space for the respective singularities is $\{Re(E),Im(E)\}$ for scalar field $E$, where $E$ corresponds to a complex scalar field for phase singularities (such as the complex amplitude of a single polarization state) and the Stokes field $\{S_1,S_2\}$ transverse C loci. For $D_{condition}=3$, the spin defect topological charge is the degree of continuous mapping between a thin spherical shell in Cartesian space $\{x,y,z\}$ and the corresponding surface in SAM space $\{\sigma_x,\sigma_y,\sigma_z\}$ (Figure \ref{fig4}c, bottom right). See the Supplementary Material for details on the topological charge of C and V singularities. 

We add here that a similar method of accumulating the polarization azimuth on a closed loop yields an integer valued quantity (sometimes called the Poincar\'e-Hopf index) for the conventional definition of a V-point, which is a transverse field zero $E_{x,y}=0$ locally surrounded by strictly linear polarization \cite{2001_Freund_OptComm,2002_Freund_OptComm}. While this quantity is topological in the sense that it is invariant under continuous deformations of the integrating curve that do not intersect the singularity, this structure is not topologically protected. This means that any perturbation in the conditions, for example, through the addition of stray light, will destroy the field zero\cite{2021_Liu_Kivshar_Nanophoton}. Even linearly polarized stray light will destroy the V-point singularity since it is only surrounded by all linear polarization azimuths, but not all linear polarization complex states. This lack of topological protection arises from the mismatch between the large condition dimension (four) and parameter dimension (two for a plane). The existence of a topological charge does not guarantee topological protection; only topological charges in the form of Equation \ref{eqn:generalcharge} with $D_{parameter}\geq D_{condition}$ are protected.

\subsection{Ensuring stability using the Field gradient}

 For well-behaved and analytic mappings, as would be the case in most physical fields, the language of topology can be simplified and we can study the \textit{Taylor expansion} of the fields in the vicinity of the singularity. Let the vector of condition values be $\bm{C}\in \mathbb{R}^{D_{condition}}$ and a position in parameter space be $\bm{a}\in \mathbb{R}^{D_{parameter}}$. Suppose there is a singularity at $\bm{a}_0$, which means that $\bm{C}(\bm{a}_0) = \bm{0}$. Then the multivariate Taylor expansion about $\bm{a}_0$ is, to leading linear order:
\begin{gather}
    \bm{C}(\bm{a}) =  \overline{J}(\bm{a}_0)(\bm{a}-\bm{a}_0) +  O\left(|\bm{a}-\bm{a}_0|^2\right)
\end{gather}

where $\overline{J}$ is a $[D_{condition}\times D_{parameter}]$ real-valued Jacobian matrix of first derivatives: $\left(\overline{J}_{ij}\right) = \frac{\partial C_i}{\partial x_j}$, $C_i$ is the $i$th condition value, and $x_j$ is the $j$th parameter coordinate. This involves calculating the Taylor expansion to the first order and examining the first gradient term, which is the derivative in 1D and the Jacobian matrix in higher dimensions. If $D_{condition}=D_{parameter}$ and the determinant of this Jacobian matrix is non-zero, the singularity possesses a charge of $m=sgn(det(\overline{J}))=\pm 1$ and is protected\cite{2000_Berry_Dennis_PRSLA,2023_Spaegele_Capasso_SciAdv}. If $det(\overline{J})$ is zero, the singularity has trivial topology ($m=0$) or is of a higher order ($|m|>1$) and is not protected against perturbations.

This field gradient methodology circumvents the complexity inherent in the integration process, especially for high-dimensional spaces. The approach is, however, not suitable for the construction and evaluation of non-generic higher-order singularities with $|m|>1$, for which the earlier high-dimensional integral or a higher order Taylor expansion will be necessary. 

In summary, the design of generic singularities requires that their dimension is equal to the difference between the dimensions of the parameter and condition spaces, and that the topological charge is $\pm 1$. The charge can be calculated either through an integral or a derivative. As derivatives tend to be more computationally efficient, they are more practical in most instances.

\section{Designing singularities}

We now turn to the problem of designing and actually realizing optical singularities. There are two main approaches to optical device design: forward and inverse design.  

\subsection{Forward design}

Forward design refers to the composition of well-understood components to build a larger device. These components can be physical elements, such as individual nanostructures or meta-elements, or they can be field decompositions where each field component is understood and can be realized directly. Forward design works well when the desired optical field distribution over the domain is known, as, for instance, in the case of a simple phase singularity with analytical azimuthal dependence $e^{i\theta}$. 

The desired field can then be decomposed into a set of simpler elements that can be individually realized and later re-composed to produce a composite field pattern. The field can be decomposed into well-understood field profiles \cite{2001_Berry_Dennis_JPhysA} or directly decomposed using any solution of the paraxial or non-paraxial Helmholtz equation, such as plane waves \cite{2021_Wang_Fan_Optica}, Bessel beams\cite{2001_Berry_Dennis_JPhysA,2016_Dorrah_Mojahedi_PhysRevA}, or Laguerre-Gaussian (LG) beams\cite{2010_Dennis_Padgett_NatPhys}. For fields that depend on the polarization transformation performed by an optical device, the each basis weight can be replaced by a Jones matrix \cite{2021_Dorrah_Capasso_NatComm}. Scalar fields can also be reconstructed using iterative (\eg computer-generated holography) or non-iterative analytic processes \cite{2020_Aborahama_Mojahedi_OptExp}. 

Those basis components can be individually created with optical components and then superposed (\eg using multiple SLMs, waveplates\cite{2013_Kumar_Viswanathan_JOpt}, q-plates\cite{2012_Cardano_Santamato_ApplOpt}, beamsplitters, mode converters, polarizers, or photonic crystals). Alternatively, the design process can be streamlined using metasurfaces by superposing these optical functions and implementing them on a single optical device. This is achieved by superposing the propagated decomposition components on the metasurface plane and implementing them using standard unit-cell approaches\cite{2021_Dorrah_Capasso_NatComm}. Metasurfaces are especially suited to realize polarization-dependent behavior, since the metasurface unit-cell can be made to exhibit spatially-varying polarization transformations and structural birefringence using simple nanostructures. 

Yet, it is non-trivial to identify the suitable optical field in the first place. While some singular structures can be identified with an analytic closed form (\eg some knots \cite{2010_Dennis_Padgett_NatPhys} and spin defect singularities \cite{2021_Wang_Fan_Optica}), in general, the singular behavior of a field does not determine its wider global behavior. This means that there are many fields that can realize the same singular behavior, although these fields are not equally useful, easily accessible, or desirable. It is often difficult to know \textit{a priori} what the optimal field distribution would be given a set of singular characteristics. Even of the field distribution is known, is can also be challenging to select the appropriate modes and their weights\cite{2009_Dennis_Padgett_ProgOpt}. In such circumstances, and in cases where the design parameters are non-intuitively connected with the resultant fields, it is useful to take an inverse design approach. 

\subsection{Inverse design}
Photonic inverse design involves expressing the system as an optimization problem over the realizable parameter space of the optical device, so that the desired fields or optical behaviors are achieved through a maximization or minimization of one or more objective function values \cite{2019_Campbell_Werner_OptMaterExp,2022_Li_Capasso_ACSPhoton}. The tunable parameters in the optimization can be the properties of a wavefront manipulating device, such as the pixel states on a spatial light modulator or the meta-element shapes on a metasurface. Inverse design is ideal for nanofabricated metasurfaces and photonic crystals, given its immense design space and range of controllable optical degrees of freedom.  In this section, we will focus on two recently-introduced design methods for singularity engineering, dividing our attention between engineering topologically protected generic singularities and non-generic singularities. 

For generic singularities, inverse design can be used to improve any step of the aforementioned forward design procedure. For instance, after selecting a set of basis waves capable of generating the desired topology, inverse design can be used to refine the coefficients of wave decomposition to maximize structural stability against fabrication imperfections and misalignment, using the field intensity as the objective function  \cite{2010_Dennis_Padgett_NatPhys}. In engineering high-dimensional singularities, one can use the determinant of the Jacobian ($\text{det}(\overline{J})$) as the objective function to ensure topological protection and maximize it to further improve the stability \cite{2023_Spaegele_Capasso_SciAdv}. This has the effect of minimizing the displacement of the singularity in parameter space upon perturbation, as the extent of movement in parameter space corresponds to the reciprocal of the Jacobian determinant. Further control over the field distribution surrounding of the singularity can be exerted by inverse-designing the Jacobian itself. For instance, by including elements of the singular value decomposition (SVD) of the Jacobian in the objective function, one can control the orientation and skew of the singularity, as detailed in the Supplementary Information of Spaegele et al\cite{2023_Spaegele_Capasso_SciAdv}.

Furthermore, when the mathematical decomposition is known, inverse design can serve as the bridge to connect the shape of the nanophotonic device, such as a metasurface or photonic crystal, to the desired transmission function needed to realize the singularity. This is typically known as topological optimization; a variant of which was implemented by Wang and Fan to fine-tune arbitrarily-tilted spatiotemporal optical vortices through constrained optimization\cite{2021_Wang_Fan_Optica}. 

However, the hardest components of inverse design remain the choice of an appropriate quantitative objective function and picking a good starting point for optimization. Forward design can be useful in identifying a good initial guess, which would reduce the number of expensive optimization steps required and increase the likelihood of achieving high-performance maxima. 

Beyond generic singularities, non-generic singularities do not exist in a random wavefield since they can be destroyed by arbitrarily small condition perturbations. Although they also do not physically exist in a strict mathematical sense in view of their fragility, they can be made to behave identically to the mathematical idealization depending on the practical purpose. Such strategies involve precise alignment of the wavefront shaping optical components, the elimination of environmental stray light, or simply selectively applying them to situations which preserve the requisite system symmetry. For example, $l=2$ vortices were employed in optical vortex grating coronagraphs due to their superior optical filtering performance over $l=1$ and $l=3$ vortices in imaging the vicinity of bright stars \cite{2005_Foo_Palacios_OpticsInfobase}, an application which preserves the circular symmetry of the higher order vortex configuration. 

\begin{figure}
\centering
\includegraphics[width=17 cm]{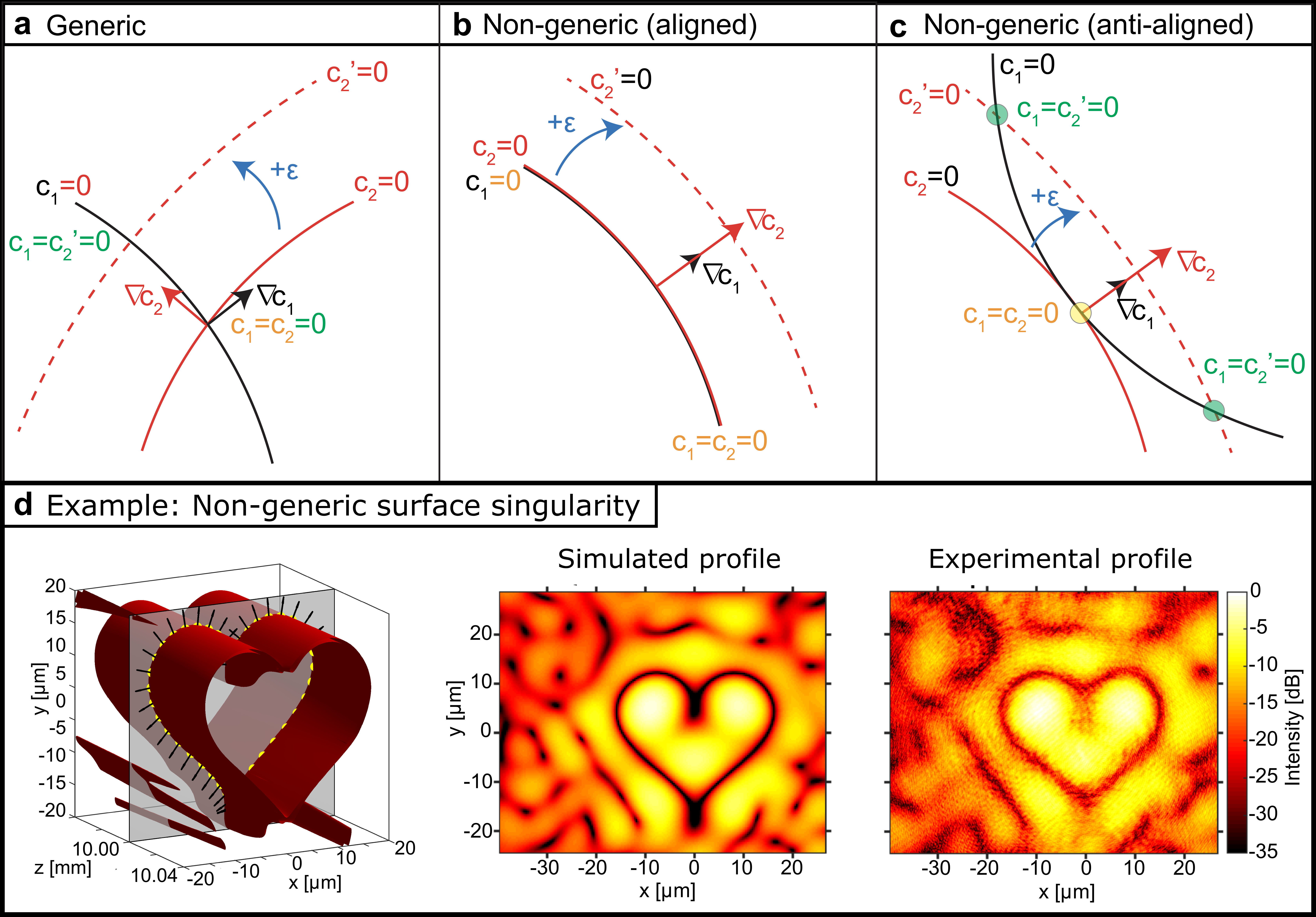}
\caption[]{
\label{fig5} Generic and non-generic intersections, depicted using condition value isolines on a 2D plane. \textbf{a} Two condition isolines intersect generically at a point, which means that their gradient vectors would be more likely to point in different directions at the singularity intersection. Gradient vectors point in the direction of maximum increase and are normal to the isoline. \textbf{b-c} When the gradient vectors of two isolines point in the same direction at the singularity intersection, their respective isolines can be aligned (\textbf{b}) or anti-aligned (\textbf{c}). Note that the local curvature of the isoline is not constrained by the direction of the gradient vector. Aligned isolines approximate an intersection along an extended locus, producing a higher-dimensional singularity locus. Anti-aligned isolines are useful in producing point phase singularities in 3D space if the condition zero level sets can be made to separate away from this 2D plane. \textbf{d} An example of a non-generic aligned surface singularity with a heart-shaped cross-section, obtained by aligning the condition gradient vectors (black arrows) along a heart-shaped contour (yellow dots). Middle: Numerically simulated and Right: experimentally-measured intensity profiles in the plane $z=0$. Images adapted with permission from S.W.D. Lim et al\cite{2021_Lim_Capasso_NatComms}, under the Creative Commons CC-BY license.}
\end{figure}

Non-generic singularities that have shape dimensions that differ from $D_{parameter}-D_{condition}$ (as would be required for generic singularities) can also be realized with inverse design. The points which satisfy each of the singularity conditions individually are $(D_{parameter}-1)$ dimensional sets: lines in a 2D parameter space (like the zero condition iso-lines in Figure \ref{fig1}a-d), surfaces in 3D parameter space (like in Figure \ref{fig1}e), and so on. The singularity exists at the intersection of those lines/surfaces/level sets. The gradient vector of a scalar field points in the direction of maximum increase of that scalar field and is thus orthogonal (in a Euclidean sense) to the level set through that point (Figure \ref{fig5}a).

While surfaces intersect generically along a line, thus having gradient vectors that point in different directions in parameter space (Figure \ref{fig5}a shows a cross-sectional cut), they can also intersect in a degenerate, non-generic manner (Figure \ref{fig5}b-c). Depending on the local curvature of the zero level sets at the intersection, the intersection locus can be aligned (Figure \ref{fig5}b), as when the level set curvature orientations match, or anti-aligned (Figure \ref{fig5}c), as when they are oriented oppositely. This is independent of the pointing direction of the gradient vector. The aligned case allows the singularity to be extended in the tangent plane of the intersection, which the anti-aligned case does not, hence the former is desirable in engineering a region of extended overlap, such as a sheet singularity \cite{2021_Lim_Capasso_NatComms} (Figure \ref{fig5}d). The anti-aligned case is useful in engineering an isolated dark point surrounded by light \cite{2023_Lim_Capasso_NatComm}. The two surfaces can also be equal so that their intersection locus is a 2D membrane; this is a 2D sheet singularity \cite{2021_Lim_Capasso_NatComms}. These degenerate intersections exhibit the fragility of the intersection --- a slight misalignment of these surfaces due to field perturbations will cause them to intersect generically along a line or not at all. 

A straightforward way to align such condition surfaces at a position is to treat the design problem as an optimization and maximize the alignment of the level set shapes at the desired singularity position. Lim et al introduced the phase gradient magnitude as one optimization objective function component, proposing that point and sheet singularities can be produced by maximizing the phase gradient magnitude along specified directions \cite{2021_Lim_Capasso_NatComms,2023_Lim_Capasso_NatComm}. The phase gradient $\nabla\phi$ is the spatial gradient of the field complex phase $\phi$. A complex scalar field can be written in polar form as $E(\bm{r}) = A(\bm{r})\exp\left[i\phi(\bm{r})\right]$, allowing the phase gradient to be written as a linear combination of the gradients of the real and imaginary field components $\nabla Re(E),\nabla Im(E)$ (Equation \ref{eqn:phasegrad}). 
\begin{gather}
\nabla\phi = Im\left(\frac{\nabla E}{E}\right) =  \frac{Re(E)}{Re(E)^2+Im(E)^2}\nabla Im(E) - \frac{Im(E)}{Re(E)^2+Im(E)^2} \nabla Re(E) \label{eqn:phasegrad}
\end{gather}

Maximizing $\nabla\phi\cdot \hat{n}$ along a particular direction $\hat{n}$ has the dual effect of maximizing the field gradients $\nabla Re(E),\nabla Im(E)$ across the singularity position and minimizing the field magnitude $|E|^2 = Re(E)^2 + Im(E)^2$. The latter enforces singular behavior (\ie moving the zero level sets to that position) while the former improves the spatial confinement of the singularity along $\hat{n}$. Tighter spatial confinement is useful for optical trapping applications. 

This field gradient optimization method can be extended to other singularity configurations by replacing the field gradients $\nabla Re(E),\nabla Im(E)$ with condition value gradients $\{\nabla c_i\}$, and by replacing the field intensity $|E|^2$ with a (possibly weighted) sum over condition magnitudes $\sum_i |c_i|^2$. 

\section{Applications of engineered singularities}

\begin{figure}
\centering
\includegraphics[width=12 cm]{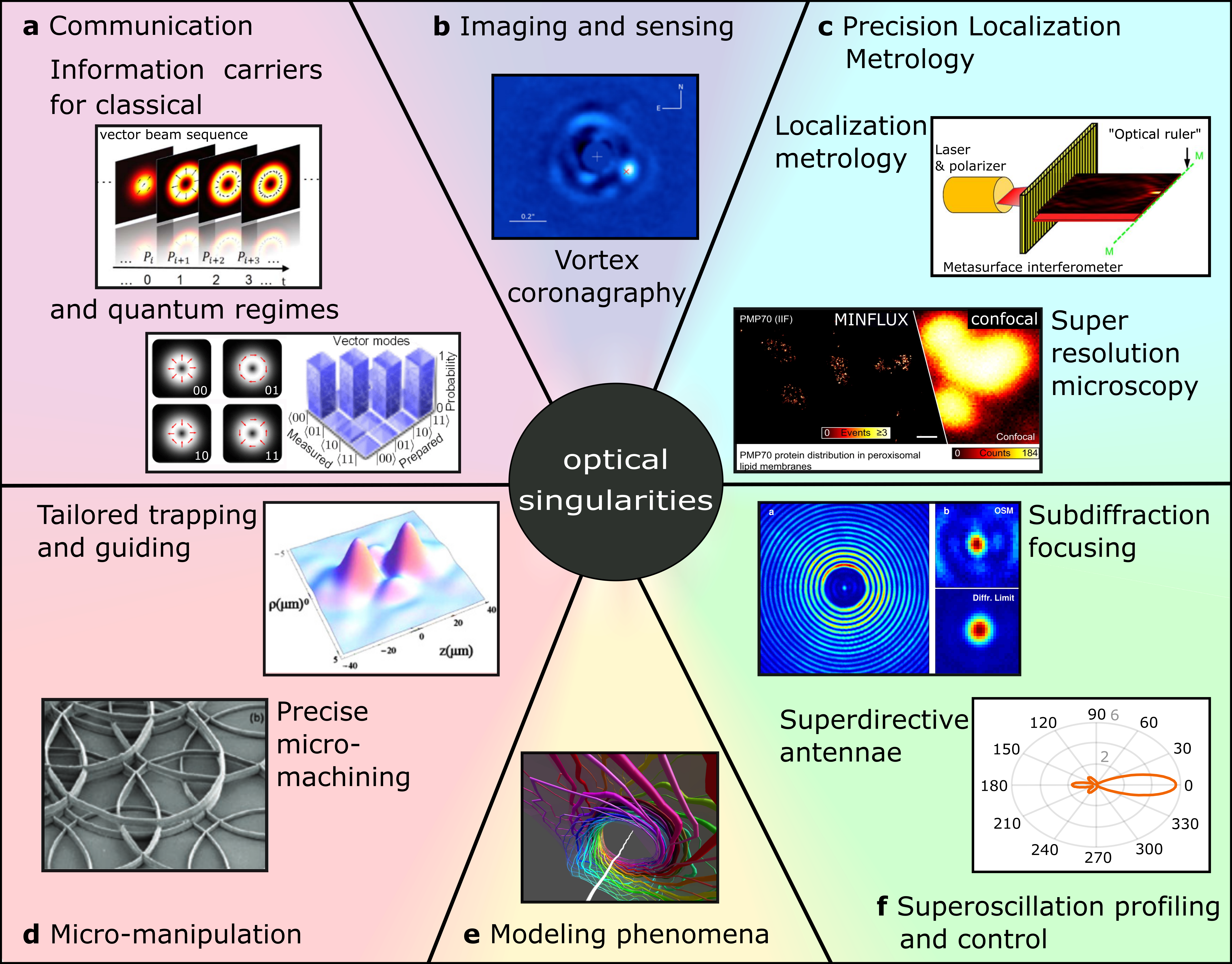}
\caption[]{Exemplary applications of optical singularities. \textbf{a} Modes with different topological charges can be used as a basis to multiplex and de-multiplex information in optical communication systems, which can increase the information bandwidth for classical communication (top) or enhance the security of quantum communication (bottom). Upper image used with permission of Chinese Laser Press from J. Wang, Advances in communications using optical vortices, Photonics Research 4, 5, B14–B28\cite{2016_Wang_PhotRes}. Lower image used with permission of The Institute of Electrical and Electronics Engineers, Incorporated (IEEE) from B. Ndagano, I. Nape, M.A. Cox, C. Rosales-Guzman, and A. Forbes, Creation and detection of vector vortex modes for classical and quantum communication. J. Light. Technol. 36, 292–301 (2018)\cite{2018_Ndagano_Forbes_JLightwaveTech}; permission conveyed through Copyright Clearance Center, Inc. \textbf{b} Dark singularities can act as spatial filters in sensing and imaging, suppressing unwanted detection regions; for example, they are utilized in vortex coronagraphy to suppress on-axis light, enhancing the detection of lower intensity off-axis sources. Image from E. Serabyn et al\cite{2017_Serabyn_Wertz_AstrophysJ}, \copyright AAS. Reproduced with permission. \textbf{c} The tight localization and high-spatial-frequency gradients encompassing optical singularities render them excellent candidates for precise localization metrology, where they can be used, for instance, as optical rulers (top) or structured beams in super-resolution microscopy techniques like MINFLUX (bottom). The latter significantly enhances the spatial resolution of fluorescence microscopy by localizing fluorescent molecules at the zero-intensity points of a pump beam. Upper image from G.H. Yuan and N.I. Zheludev, Detecting nanometric displacements with optical ruler metrology, Science 364, 771–775 (2019)\cite{2019_Yuan_Zheludev_Science}. Reprinted with permission from AAAS. Lower image from R. Schmidt et al\cite{2021_Schmidt_Hell_NatComm}, used under the Creative Commons CC BY license. \textbf{d} The tight localization of dark singularities and the ability to design their (polarization-dependent) environment makes them an ideal candidate, \eg for advanced trapping and manipulation of particles and high precision photon-based micro-machining. Upper image used with permission from P. Xu, X. He, J. Wang and M. Zhan\cite{2010_Xu_Zhan_OptLett}. \copyright Optica Publishing Group. Lower image used with permission of John Wiley and Sons, from M. Duocastella and C. Arnold, Bessel and annular beams for materials processing, Laser \& Photonics Rev. 6, 607–621 (2012)\cite{2012_Duocastella_Arnold_LaserPhotRev}; permission conveyed through Copyright Clearance Center, Inc. \textbf{e} Optical singularities enable modeling of complex phenomena across various domains of physics by serving as scalable, controllable laboratory-scale proxies, facilitating insights into high-dimensional geometries and topological states. Image copyright Danica Sugic and Mark Dennis, University of Birmingham. The image represents the work published in Sugic et al, Particle-like Topologies in Light, Nature Communications \textbf{12}, 6785 (2021)\cite{2021_Sugic_Dennis_NatComm}, \href{https://www.nature.com/articles/s41467-021-26171-5}{https://www.nature.com/articles/s41467-021-26171-5}. \textbf{f} Optical singularities enable the creation of superoscillations, which have been used to generate focal spots smaller than the diffraction limit (top) or highly directed signals (bottom). Upper image used from A. Wong and G. Eleftheriades\cite{2013_Wong_Eleftheriades_SciRep} under the Creative Commons Attribution-NonCommercial-NoDerivs 3.0 Unported License. Lower image adapted with permission from W. Chen, J. Fu, B. Lv, and Q. Wu\cite{2020_Chen_Wu_ApplOpt}. \copyright Optica Publishing Group.
\label{fig6}}
\end{figure}

We now turn to categorize the key application domains for engineered optical singularities, which are summarized in Figure \ref{fig6}. 

\subsection{Orbital angular momentum and Communication}\label{OAM}

Similar to the Earth, which possesses angular momentum due to both its spin on its axis and its orbit around the Sun, optical fields also have spin and orbital angular momentum. The spin angular momentum of light results from its transverse polarization --— the time-dependent rotation of the electric field vector as it outlines the polarization ellipse --- reaching its extremal values with circularly polarized light. On the contrary, orbital angular momentum (OAM) is linked to the spatial configuration of the wavefront. It was identified relatively recently, in 1992, by L. Allen et al\cite{1992_Allen_Woerdman_PhysRevA}. 

The vortex-like, helical phasefront of a cylindrically-symmetric OAM beam has become synonymous with OAM in general. Since such OAM beams possess an on-axis linear phase singularity around which the Poynting vector swirls, it is easy to associate OAM directly with optical singularities. Furthermore, the equivalence of the topological charge $m$ of Laguerre-Gaussian beams and the OAM per photon $m\cdot\hbar$ of such beams might suggest that the topological charge is equivalent to the OAM per photon of optical beams.

Regrettably, this neat correspondence between OAM and optical singularities fails for optical beams other than  LG modes. OAM is a property of any beam with a wavefront tilt with respect to the optical axis and does not require the presence of a phase singularity. In 1997, J. Courtial observed that an elliptical Gaussian beam without a phase singularity can have arbitrary OAM values per photon, as high as $10000\cdot\hbar$\cite{1997_Courtial_Padgett_OptComm}. Although the total angular momentum and orbital angular momentum of a beam are conserved during propagation in free space, optical singularities can form and annihilate upon propagation, so that the total topological charge of the beam is not conserved during propagation\cite{1997_Soskin_Heckenberg_PhysRevA}. More recently, Berry constructed a beam with axial singularities and a non-zero total topological charge but zero total OAM\cite{2022_Berry_Liu_JPhysA}. Undoubtedly, OAM is a powerful concept with numerous applications. However, as it exists independently of optical singularities, we will focus here on applications which require the properties of optical singularities. A thorough exploration of OAM physics and applications can be found in the reviews of Allen et al\cite{1999_Allen_Babiker_OAMLight,2011_Allen_Padgett_Book} and Yao et al\cite{2011_Yao_Padgett_AdvOptPhoton}. 

One key exception is the use of structured light containing OAM for information multiplexing: to transmit classical information \cite{2004_Gibson_FrankeArnold_OptExp,2012_Wang_Willner_NatPhoton,2013_Bozinovic_Ramachandran_Science,2016_Wang_PhotRes,2020_Larocque_Karimi_NatComm}, perform simple arithmetic \cite{1997_Berzanskis_Stabinis_OptComm,1999_Freund_OptComm} and even be involved in quantum communication\cite{2018_Ndagano_Forbes_JLightwaveTech}. The property of a set of modes by which information can be multiplexed and de-multiplexed --- orthogonality --- requires phase singularities to be present in the mode profiles. Orthogonality is the property that the inner product between two different eigenmodes is zero (and strictly positive when evaluated for the same eigenmode), and this allows a linear combination of eigenmodes to be unambiguously decomposed into its constituent eigenmodes by taking its inner product with each member of an eigenmode set. For modes that occupy the same space (\ie not spatially separated like in uncoupled many-core fibers), orthogonality requires that all but one of the modes (usually the lowest-order mode) have field zero-crossings, which are zero lines or surfaces in 3D. To see this, consider the example of the inner product for two propagating scalar modes (indexed $a$ and $b$) defined over a transverse $xy$ region $S$ (Equation \ref{eqn:innerproduct}):
\begin{gather}
    \langle \phi_a |\phi_b \rangle = \iint_S w(\bm{r})\phi_a(\bm{r})\phi_b(\bm{r})d^2\bm{r}\label{eqn:innerproduct}
\end{gather}

$w(\bm{r})$ is a non-negative weighting function. It is non-negative because of the requirement that $\langle \phi_a|\phi_a\rangle > 0$ for all eigenmodes $\{a\}$. We want to show that there is at most one eigenmode in an orthogonal basis that has no sign changes and thus no phase singularities. Assume the contrary; let there be more than one eigenmode in an orthogonal basis with no sign changes. Due to the continuity of physically-realizable modes, each of these eigenmodes will have the same sign over $S$, and their product will also have the same sign over $S$. This means that $\langle \phi_a |\phi_b \rangle$ cannot be zero except for the trivial case where one or more of the eigenmodes is zero everywhere. Thus these eigenmodes cannot be part of an orthogonal basis, proving by contradiction that there is at most one eigenmode in an orthogonal set that has no sign changes; all other eigenmodes must have phase singularities. These nodal singularities have not been explicitly engineered for communication purposes, which raises the intriguing possibility that the singularities in orthogonal eigenmodes can be deterministically positioned at known obstacle locations to communicate around them with minimal scattering loss. 

\subsection{Imaging and Sensing}\label{sensing}

In 1976, within the context of a bright singularity for ray optics known as caustics, M.V. Berry conjectured that structurally stable, generic singularities could serve as a form of ``skeleton'' for the wavefield. Around these ``skeletons'', approximations to the global wave behavior could be constructed\cite{1976_Berry_AdvPhys}. Since optical phase singularities are generic properties of wavefields, such singular ``skeletons'' can serve as markers for tracking the evolution of a field\cite{1992_Zuravlev_Kravtsov_ZETF,1998_Aksenov_Tikhomirova_ApplOpt}. The positions of intensity zeros on a 2D transverse plane have been used to construct an improved starting guess for phase retrieval algorithms in image reconstruction\cite{1994_Wackerman_Yagle_JOSAA,1994_Chen_Fiddy_JOSAA}. They can even perform blind deconvolution when the number of such zeros is approximately equal to the number of samples used to represent an image function\cite{1996_Chen_Pommet_JOSAA}. Including the vortex-like phase profiles of the singularities also allows for more efficient phase correction when applied to adaptive optics\cite{2001_Banakh_Falits_SPIE} and can even be used to deduce the refractive index of a medium across a wide dynamic range\cite{2018_Dorrah_Mojahedi_LSA}.

A major contribution of optical singularities to imaging was the realization that the singularities produced by optical devices reflect the coherence and tilt angle of the illuminating light field. The formation of intensity zeros from a flat phase plate, which involves destructive interference, requires coherent illumination and normal incidence. Incoherent illumination and off-axis incidence cannot undergo complete destructive interference and thus cannot produce intensity zeros. The first spatial filter based on vortex phase profiles was realized in 1992 by Khonina et al\cite{1992_Khonina_Uspleniev_JModOpt}. In 2001, G. Swartzlander Jr. proposed the vortex beam phase mask as a spatial filter that can discriminate between on-axis coherent radiation and other incoherent or off-axis contributions\cite{2001_Swartzlander_OptLett}. Only on-axis coherent radiation can form the dark spot at the center of a vortex beam and would thus be suppressed in the imaging plane relative to that of incoherent or off-axis illumination sources. This principle was used to detect forward-scattered light from a colloidal solution by nulling the strong unscattered beam\cite{2002_Palacios_Swartzlander_PRL}. It also formed the working principle of the optical \textit{vortex coronagraph}\cite{2005_Foo_Palacios_OpticsInfobase}. Vortex coronagraphy operates by suppressing the strong on-axis contribution of a bright light source, such as an on-axis star in astronomical imaging, so that peripheral objects, which would otherwise be overshadowed, can be detected. Numerical predictions from Foo et al indicated that vortices with a topological charge magnitude of $2$ provided better contrast between an unsuppressed off-axis source and the suppressed on-axis source compared to charges of $1$ and $3$\cite{2005_Foo_Palacios_OpticsInfobase}. Mawet et al later identified that for coronagraphs made from subwavelength phase gratings, only the charge $2$ mask had the correct geometrical symmetry to produce quasi-achromatic behavior\cite{2005_Mawet_Surdej_AstrophysJ}. A vortex coronagraph that employed a charge $2$ 3D surface-relief vortex phase shifter was later demonstrated in an astronomical imaging system incorporating an 8-inch refracting telescope and adaptive optics. It successfully resolved a binary star separated from its brighter parent by an angular displacement of $1.9\lambda/D$\cite{2008_Swartzlander_Wilson_OptExp}. More recently, in 2017, an annular groove phase mask (subwavelength-grating phase mask) infrared vortex coronagraph was incorporated into the 10 m primary mirror Keck II ground-based telescope. It was used to detect a brown dwarf (HIP 79124 B) that was separated from its brighter companion star by a mere 186.5 milli-arcseconds ($5.2\times 10^{-5}$ degrees)\cite{2017_Serabyn_Wertz_AstrophysJ}.

\subsection{Precision Localization Metrology}

Since large phase gradients are tightly localized to the region around the phase singularity, they are narrow, high-spatial-frequency features of a field that can be tracked to resolve minute field displacements. In 2019, G.H. Yuan and Zheludev exploited this behavior to create an optical ruler for measuring spatial displacements with sub-nanometer resolution\cite{2019_Yuan_Zheludev_Science}. They demonstrated a self-referenced interferometric technique to measure the phase profile of a phase singularity, showing that minute field displacements (approximately $\lambda/800$) could be detected by tracking the narrow, high phase gradient features of the field. In contrast, tracking bright peaks resulted in much poorer spatial resolution (approximately $\lambda/3$) due to the finite wavelength-scale width of these peaks.

Recent further work by Liu et al from Zheludev's group has employed machine-learning techniques on scattered light fields to reconstruct nanowire displacements with better than 92 pm (0.092 nm, or approximately $\lambda/5300$) precision\cite{2023_Liu_Zheludev_NatMater}. This technique relies on the existence of large momentum transfers in high phase gradient regions: by placing a nanowire in a structured light field containing phase singularities and high phase gradient regions, the light field scattered off the nanowire becomes strongly dependent on the nanowire's precise position within the light field. A trained neural network can then estimate the nanowire displacements based on the observed scattered field patterns.

\begin{figure}
\centering
\includegraphics[width=10 cm]{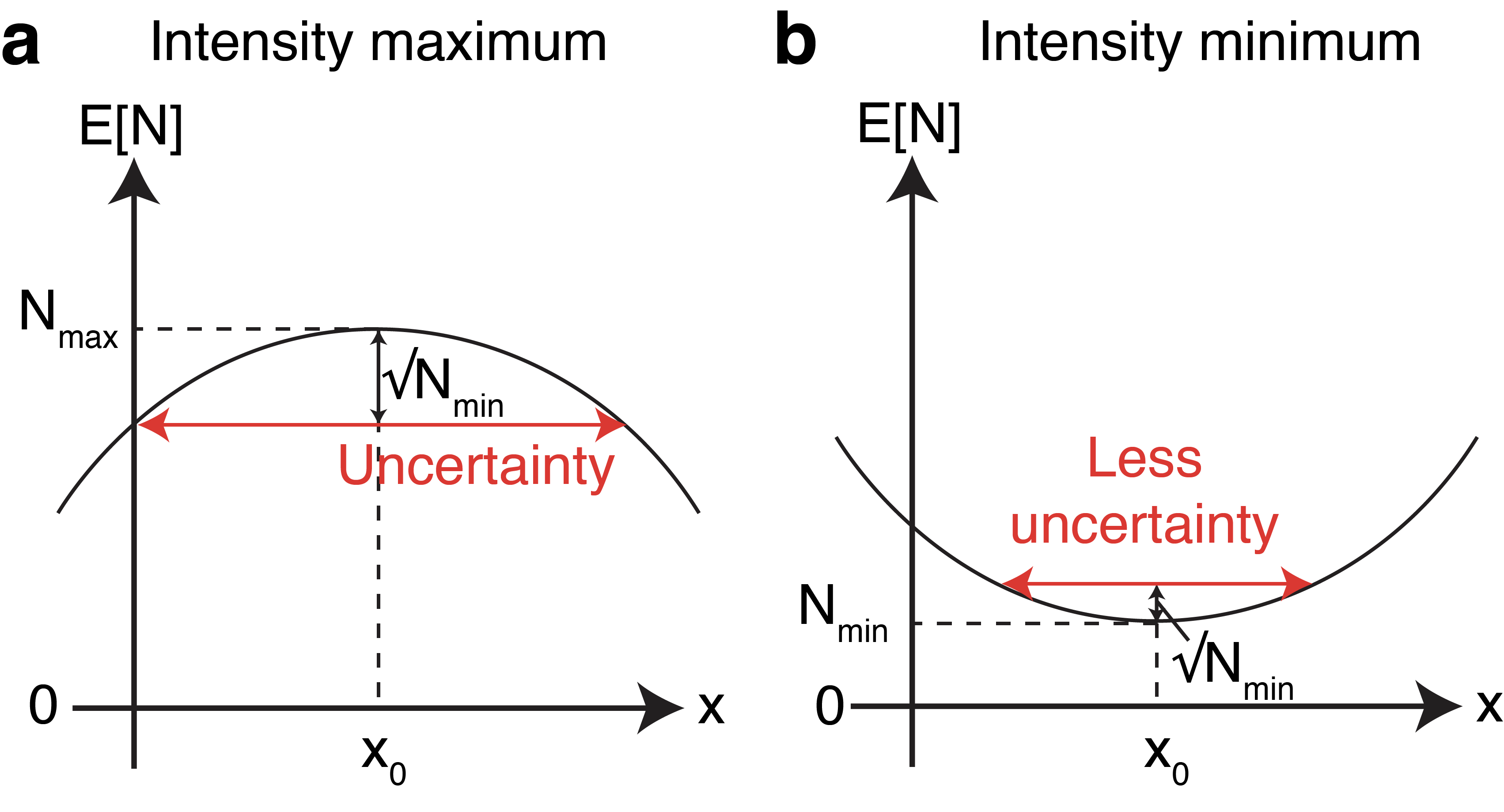}
\caption[]{Model for the localization precision of an intensity maximum versus intensity zero. The uncertainty in a localization measurement is the range in $x$ position that is consistent with a signal within one standard deviation $\sqrt{N}$ of the feature signal strength $N$. \textbf{a} An intensity maximum has a higher central intensity compared to that of the intensity minimum $N_{max}\gg N_{min}$. Thus, it has a large intensity standard deviation and a large positional uncertainty. \textbf{b} An intensity minimum has a small minimum intensity and thus has a reduced localization error. For a phase singularity, $N_{min}$ can be made near zero. Image adapted from J. Xiao and T. Ha, Flipping nanoscopy on its head, Science 355, 582–584 (2017)\cite{2017_Xiao_Ha_Science}. Reprinted with permission from AAAS.}
\label{fig7}
\end{figure}

Even for applications that are only sensitive to field intensities and lack phase resolution, the measurement of field zeros can be more precise than that of field maxima. The underlying reason lies in the shot noise statistics, which offer more favorable scaling behavior with photon counts when one uses field minima. In 1949, Hans Wolter chose to track the displacements of beams using their intensity minima, rather than maxima, in total internal reflection experiments\cite{1950_Wolter_Naturf,1950_Wolter_ZNaturforsch,2009_Wolter_JOA}. Shot noise refers to the fluctuations in field intensities due to the quantized nature of light. Photons in an optical signal can be treated using Poisson statistics, where each photon incidence is considered an independent event. If the expectation value of the number of events $N$ is $E[N]$, $E[N]$ is proportional to the expectation value of the light intensity, and the standard deviation in measurements of $N$ is $\sqrt{E[N]}$. Xiao and Ha provide a simple model connecting shot noise errors to localization precision, considering two positions on a signal to be spatially resolved if the difference in signal intensity is larger than the intensity standard deviation\cite{2017_Xiao_Ha_Science} {(Figure \ref{fig7}). If the intensity maximum has a signal expectation value of $N_{max}\gg 1$, the signal standard deviation $\sqrt{N_{max}}$ is large, indicating that a broad range of points have signal intensities that are consistent with that of the true maximum, resulting in a large localization error {(Figure \ref{fig7}a). In contrast, for an intensity minimum with the same local intensity curvature, since the minimum intensity $N_{min}$ is much smaller than $N_{max}$, the corresponding signal standard deviation $\sqrt{N_{min}}$ is small, leading to tight localization (Figure \ref{fig7}b). In fact, if the intensity minimum corresponds to that of a phase singularity, then $N_{min}$ can be near zero and is only bound from below by the background signal from other environmental sources.

The superior localization of dark spots compared to diffraction-limited bright focal spots was exploited in the Nobel Prize-winning superresolution microscopy work in STED. In STED, a fluorescent particle is illuminated with a superposition of two beams: one focused pump beam that causes the particle to fluoresce, and a focused depletion beam with an isolated dark spot that is smaller than the pump beam. The particle's fluorescence will be suppressed if it is illuminated by the depletion beam. When the dark spot of the depletion beam overlaps with the larger bright focal spot from the pump beam, the fluorescent particle can only fluoresce if it is in the smaller dark spot of the depletion beam. This reduces the effective axial and transverse extents of the pump beam spot, increasing the spatial resolution of the fluorescent measurement as the two-color focal spot is scanned around the region of interest\cite{1994_Hell_Wichmann_OptLett,2007_Willig_Hell_NatMethods}. STED microscopy has achieved 3D spatial resolutions down to approximately $\lambda/10$. Further resolution improvements in STED have been limited by the localization precision of the fluorescent emission: for a Gaussian spot with a width of $\sigma(\lambda)$, which scales linearly with the emission wavelength, the localization precision of the Gaussian centroid is $\sigma(\lambda)/\sqrt{N}$, where $N$ is the number of photons collected. Therefore, high spatial resolution STED captures include long integration times to increase $N$, during which the particle may photobleach or transition to other non-emitting states\cite{2017_Balzarotti_Hell_Science}.

In 2017, Balzarotti et al from Stefan Hell's group introduced the MINFLUX superresolution microscopy technique\cite{2017_Balzarotti_Hell_Science}, which achieved an order of magnitude improvement in spatial and temporal resolution\cite{2018_Eilers_Hell_PNAS,2021_Schmidt_Hell_NatComm,2023_Deguchi_Ries_Science}. Instead of localizing fluorescent molecules at the maximum of a pump beam, the key innovation of MINFLUX was to localize these molecules at the zero-intensity point of a pump beam. The intensity of fluorescent emission depends on the pump beam intensity at the position of the fluorescent molecule. Thus, by illuminating the particle with a scanning pump beam that contains a single isolated point phase singularity, one can identify the position of the particle by determining the beam position that minimizes the fluorescent emission. In effect, the fluorescent particle serves as a point measurement of the local pump beam intensity. With as few as four measurements at different beam positions, MINFLUX can obtain an estimate of the particle position. The localization precision for MINFLUX scales as $L/\sqrt{N}$ for one such set of measurements, where $L$ is now a wavelength-independent length that corresponds to the spatial extent sampled by the four measurements\cite{2017_Balzarotti_Hell_Science}. Since $L\ll \sigma(\lambda)$, MINFLUX attains a much higher spatial resolution than STED for the same photon flux. MINFLUX can be operated iteratively — by repeatedly ``zooming'' into the particle with another set of four measurements and reducing the measurement spatial extent each time, the spatial resolution of iterative MINFLUX improves exponentially: $e^{-N}$, rapidly shrinking with the number of photons $N$ collected\cite{2020_Gwosch_Hell_NatMethods,2021_Schmidt_Hell_NatComm}. The need for a very small number of photons for localization allows MINFLUX to be operated at very rapid timescales, and it has recently been used to identify the ``walking'' dynamics of the motor protein kinesin in living cells at 2 nm spatial resolution and millisecond temporal resolution\cite{2023_Deguchi_Ries_Science}. 

\subsection{Micro-manipulation}

The intensity minima associated with phase singularities have been employed to trap and manipulate microparticles and cold atoms. Single-beam optical traps conventionally use a tightly focused intensity maximum to stably trap high-index microparticles, as in the earliest experimental demonstrations by Ashkin et al \cite{1986_Ashkin_Chu_OptLett}. In 1995, He et al achieved trapping of absorptive and reflective particles in a doughnut (vortex) beam, and even manipulated these particles by setting them in rotational motion through the transfer of OAM \cite{1995_He_RubinszteinDunlop_JModOpt}. Soon after, in 1996, Gahagan et al employed a focused vortex beam to trap hollow glass spheres in water, exploiting the lower index of refraction of the particle compared to the surrounding medium so that the particle was repelled from high intensity regions into dark regions\cite{1996_Gahagan_Swartzlander_OptLett}. A review of micromechanical manipulation physics and techniques using OAM-bearing vortices is available by Heckenberg et al \cite{1999_Heckenberg_RubinszteinDunlop_OptVort}. 

Dark phase singularities also play an essential role in atomic physics since they can be used to trap and manipulate cold atoms using blue-detuned light. Most cold atom traps employ light that is red-detuned (\ie has a longer wavelength) from a strong dipole resonance since such red-detuned light exerts an attractive force to trap atoms in an intensity maximum. Blue-detuned light is repulsive and hence can trap cold atoms in intensity minima instead. The earliest mention that a low intensity region could be used to trap cold atoms was in 1992 by Bazhenov et al, who speculated that a hollow beam might be able to trap and guide atoms \cite{1992_Bazhenov_Vasnetsov_JModernOpt}. Such light-based structures were soon experimentally realized using vortex modes \cite{1997_Kuga_Sasada_PRL,1998_Yin_Jhe_JOSAB}. 

The tight localization of the central singularity in a vortex beam has also been used for micromachining ring-shaped structures using ultrashort pulses containing an on-axis singularity \cite{2010_Hnatovsky_Rode_OptLett,2010_Hamazaki_Omatsu_OptExp}. Hnatovsky et al also investigated the effect of various spatial polarization distributions on the ablation produced by tightly focused ultrashort pulses containing vortex modes, demonstrating that the longitudinal field had to be considered as it was able to ablate material \cite{2012_Hnatovsky_Krolikowski_OptLett}. Duocastella et al has presented a review of micromachining techniques using optical vortices \cite{2012_Duocastella_Arnold_LaserPhotRev}. 

\subsection{Modeling phenomena}

The topological structure of light around singularities can be mapped to other domains of physics, allowing these optical structures to serve as convenient, easily-manipulated, and laboratory-scale proxies for more complex phenomena. A skyrmionic hopfion, for instance, which comprises one C line threading another C line of opposite handedness, is a 3D topological soliton that has analogous realizations in solid-state, high-energy, and cosmological physics \cite{2021_Sugic_Dennis_NatComm}. Higher-order skyrmionic hopfion have also been experimentally realized \cite{2023_Ehrmanntraut_Denz_Optica}. The large number of tunable parameters, such as frequency and transverse tilts, that control the properties of light fields can also serve as synthetic dimensions, allowing the modeling of high-dimensional geometries \cite{2018_Yuan_Fan_Optica}. Topologically non-trivial light can also excite topological collective states in atomic systems, which may have implications for information transmission and quantum simulators \cite{2022_Parmee_Ruostekoski_CommsPhys,2023_Ehrmanntraut_Denz_Optica}. 

\subsection{Superoscillation profiling and control}

Superoscillations are local wave behaviors in bandlimited functions which seemingly contain Fourier components outside the given function's frequency range. Singularities have a deep connection to superoscillations --- superoscillations occur near singularities and have properties influenced by the geometric features of these singularities. Optical singularities are the control knobs for the manipulation and behavior of nearby superoscillations.

Superoscillations have their historical roots in antenna theory and intensity zeros. In 1943, S.A. Schelkunoff introduced linear transmitting antenna arrays comprising elements with subwavelength spacing that could emit signals with very strong directionality\cite{1943_Schelkunoff_BellLabs}. In 1952, Toraldo Di Francia introduced a systematic approach to constructing these ``super-gain'' antennas that emitted radiation in an extremely narrow angular range in the far-field\cite{1952_DiFrancia_IlNuovoCimento}. When applied to structuring the radial profile of 2D apertures, this \textit{Toraldo pupil synthesis} allows one to construct apertures with a far-field diffraction central lobe narrower than that of the diffraction-limited lobe size for open apertures of diameter $D$: $\Delta\theta = 1.22\lambda/D$. The Toraldo pupil synthesis procedure is summarized in the Supplementary Information, which shows how far-field intensity zeros (singularities) can be deterministically positioned in the far-field to produce sub-diffraction central lobes. 

The Toraldo pupil concept was extended to design lenses that focus light into spots smaller than the diffraction limit. These optical systems were termed ``super-resolving.'' In 1986, Hegedus and Sarafis constructed a two-zone Toraldo pupil filter for a confocal scanning system and observed that the main transverse lobe in its point spread function was halved in transverse size. However, it had an intensity 300 times lower than that of an open aperture and was surrounded by bright concentric sidebands\cite{1986_Hegedus_Sarafis_JOSAA}. 

Despite some early validation of Toraldo synthesis as super-resolving filters for focusing systems, the first experimental validation of Toraldo pupils in diffraction came five decades after di Francia's proposal, and was conducted by Mugnai, Ranfagni, and Ruggeri in the microwave regime \cite{2003_Mugnai_Ruggeri_PhysLettA,2004_Ranfagni_Ruggeri_JAP,2007_Mugnai_Ranfagni_JAP}. Activity in super-resolving optical systems based on Toraldo synthesis and more complex optimization strategies for subwavelength focal spots grew rapidly in the early 21st century\cite{2009_Huang_Zheludev_Nanolett,2014_Huang_Qiu_LaserPhotRev,2017_Yuan_Zheludev_LSA,2017_Qin_Hong_AdvMater,2018_Li_Luo_LaserPhotRev,2018_Rogers_Rogers_OptExp,2019_Gbur_Nanophoton,2019_Chen_Qiu_LSA}. These have been employed in super-resolution microscopes\cite{2012_Rogers_Zheludev_NatMater,2013_Wong_Eleftheriades_SciRep,2013_Rogers_Zheludev_JOpt,2017_Qin_Hong_AdvMater}, Raman spectroscopy\cite{2012_Kim_Stranick_OptExp}, and radio astronomy\cite{2017_Olmi_Stefani_ExpAstron}.

The theoretical basis for super-gain antennae and super-resolving pupils was provided by the mathematical concept of superoscillation. In the late 1980s, Y. Aharonov and collaborators revealed several paradoxical predictions about anomalously large values of quantum weak measurements\cite{1988_Aharonov_Vaidman_PRL} and apparent Hamiltonian energies \cite{1990_Aharonov_Vaidman_PRL}. Both of these phenomena arise from a bandlimited function oscillating faster than its non-zero Fourier components. M. Berry conducted the first extensive study on such wave behavior and named them ``superoscillations''\cite{1995_Berry_FasterThanFourier} and demonstrated that they were not strictly quantum effects\cite{1994_Berry_JPhysA}. Further mathematical properties of 1D superoscillatory functions were elucidated by A. Kempf\cite{2000_Kempf_JMathPhys} and P. Ferreira\cite{2006_Ferreira_Kempf_IEEETransSigProcess}, who provided scaling bounds for the energy, interval, and number of superoscillations. The first connections that superoscillations underlie focal spots smaller than the diffraction limit were made by F.M. Huang et al, who observed subwavelength hotspots in the near-field of nanohole arrays, and speculated that it may have been enabled by superoscillations\cite{2007_Huang_deAbajo_APL,2007_Huang_Zheludev_JOptA}. 

Superoscillations require phase singularities. The large phase gradients $\nabla\phi$ surrounding optical phase singularities are examples of superoscillations\cite{2009_Berry_Dennis_JPhysA}, and more generally superoscillations only occur in the vicinity of such phase singularities. This proximity to intensity zeros is consistent with the occurrence of superoscillations in low-intensity regions of the field where there is near-perfect destructive interference. In fact, the greater the degree of superoscillation, the closer the phase singularity and thus the lower the field intensity in the vicinity of the superoscillation. The singularities are not always obvious: in the canonical superoscillatory function\cite{2006_Berry_Popescu_JPhysA,2019_Berry_Hong_JOpt}, $f(x) = \left(\cos x + ia\sin x\right)^N, a\in\mathbb{R},N\in\mathbb{Z}$, which is defined on the real line, there are no zeros or phase singularities. However, when one plots the field in the complex plane, it becomes apparent that the superoscillatory behavior is derived from nearby charge-$N$ singularities, displaced away from the real-line in the imaginary direction (Supplementary Figure 2). The Supplementary Information provides further details about the structure of this superoscillatory function. 

Although superoscillations have only been connected to phase singularities (as opposed to polarization singularities, for instance) to the authors' knowledge, which produces their characteristic tight intensity localization, it is conceivable that a similar formalism can be extended to higher-dimensional singular features associated with intensity minima, such as transverse V and vectorial V singularities. 

\section{Conclusion and outlook}

The mathematical structure and creative applications of optical singularities belie their simple definition as electromagnetic zeros. Their behavior is reminiscent of uncertainty principles that arise from the measurement uncertainties of Fourier conjugates such as those of position-momentum and time-energy: electromagnetic waves can exhibit local properties that appear to violate their global constraints (\eg superoscillations), but only for a small region of space and time and with low probability density. Yet, judicious use of this small, sensitive region of spacetime has enabled powerful tools in imaging, microscopy, and metrology --- while mostly using the simplest phase singularities. The many empty cells in Table \ref{table2} for various dimension/co-dimension combinations indicate that the work in singular optics is far from over.

Due to the high localization precision of optical singularities, singularity-enabled applications require sensors which not only have high spatial resolution but which also have tilt (phase) sensitivity. Wavefront sensors are critical for the detection and metrology of optical singularities, especially phase singularities. One can attain deeply subwavelength localization of a optical singularity by profiling the field gradients in space since these phase gradients attain these extremely large values over a very narrow region of space that is much smaller than the wavelength-scale limit set by the diffraction of light. There are thus opportunities in the development of polarimetric versions of such high resolution wavefront sensors, which will enable full complex field tomography of light (\ie complex $E_x,E_y,E_z$ measurements in space).  This will in turn will enable tight localization of polarization singularities and the construction of other classes of high-dimensional topologically protected singularities. High frame rate singularity speckle tracking may also enable more precise non-contact measurements of minute displacements in the time-domain, enhancing techniques like laser speckle rheometry \cite{2013_Hajjarian_Nadkarni_PLoSONE,2015_Hajjarian_Nadkarni_OptLett}.

Dark singularities may thus be ideal positions for covert sensors to be located. The intensity of scattered light away from a scatterer depends on the local field intensity at the position of the scatterer. Consider, for example, a propagating electromagnetic field designed to have an approximate volumetric singularity (\ie 3D volume of very low light intensity) around a point. A scatterer placed within that position will cause minimal scattering of the local field. Information about movements and perturbations that occur within that region will also not be imprinted onto the far-field pattern of the source electromagnetic field. Such ``dead zones'' may be ideal positions to place devices that may otherwise shadow the illuminating field and give away their existence, or to place sensitive sensors that may be damaged if the illuminating field is used as an adversarial blinding beam. 

\noindent\textbf{Acknowledgments}
The authors thank Dr. A.H. Dorrah from Harvard University for valuable feedback. This material is based on work supported by the Air Force Office of Scientific Research under award numbers FA9550-21-1-0312 (F.C.) and FA9550-22-1-0243 (F.C.). S.W.D.L. is supported by A*STAR Singapore through the National Science Scholarship scheme. 

\noindent\textbf{Competing interests}
S.W.D.L. and F.C. are the inventors on a relevant provisional patent application (application number: US20230021549A1) owned by Harvard University. The authors declare no other competing interests.

\noindent\textbf{Key points}
\begin{itemize}
    \item Despite the wide range of singular optical structures described in literature, only a small set of parameters can be singular: phase, field directions, polarization ellipse parameters, and the polarization state. 
    \item All optical singularities can be described by their parameter space (which specifies the location of the singularity) and the condition space (which describes what is happening at each position in parameter space). 
    \item Singularity stability is characterized by the topological charge, which must be $\pm 1$ for stability against field or environment perturbations.
    \item Singularities can be sculpted through a range of forward and inverse design protocols. High performing inverse design objective functions may include field gradients. 
    \item Key applications of singular regions include classical and quantum communication, imaging and sensing, precision localization metrology, superoscillation profiling and control, modeling of complex physical systems, and micro-manipulation. 
\end{itemize}

\bibliography{references}

\newpage
\renewcommand{\figurename}{Supplementary Figure}

\section{Supplementary Information}
\subsection{Mathematical nomenclature} 

Throughout, we use $unbolded$ symbols (\eg $\phi,f$) to denote scalars, which can either be complex ($\mathbb{C}$) or real-valued ($\mathbb{R}$). Boldface symbols ($\bm{E},\bm{H}$) represent column vectors and vector-valued functions, and overlined symbols ($\overline{\bm{J}}$) indicate matrices and tensors.

Time-harmonic electromagnetic fields follow the convention $e^{i\omega t}$. The real and imaginary components of the complex number are written $E = E_r + iE_i$, $E_r=Re(E)$, and $E_i = Im(E)$. The intensity $I$ of a scalar field is given by $|E|^2$, and the phase $\phi$ is the four-quadrant arctangent, $\phi = \text{atan2}(E_i,E_r)$. We choose the direction of propagation when it exists to be the longitudinal/axial $z$ axis, denoting it with the ``perpendicular'' symbol, such as $E_\perp,H_\perp$. The transverse directions, $x$ and $y$, are denoted with the ``parallel'' symbol: $\bm{E}_\parallel = (E_x,E_y), \bm{H}_\parallel = (H_x,H_y), \nabla_\parallel = (\partial_x,\partial_y)$.

The transverse polarization vector $\bm{E}_\perp$ is known as the Jones vector. Transverse polarization can be parametrized in terms of the polarization ellipse on the transverse plane, which is parametrized by two variables: the azimuth $\Psi$, which is the angle that the major axis of the polarization ellipse makes with the transverse $x$-direction, and the ellipticity angle $\theta$, which determines the eccentricity and handedness (e.g., right handed or left handed, based on the temporal path of the electric field vector) of the ellipse. 

The transverse polarizations in the Jones vector can also be expressed in terms of the Stokes polarization parameters. These are four experimentally measurable intensities defined below using the left-handed convention\cite{1993_Collett_PolarizedLight}.
\begin{gather}
S_0 = I_x + I_y = |E_x|^2 + |E_y|^2 \\
S_1 = I_x - I_y = |E_x|^2 - |E_y|^2 \\
S_2 = I_{45^\circ} - I_{-45^\circ} = 2 \text{ Re}(E_xE_y^*) \\
S_3 = I_{LCP} - I_{RCP} = 2\text{ Im}(E_xE_y^*)
\end{gather}

Here, $I_j$ denotes the intensity of the polarization component projected in the $j$ direction. For fully polarized light, $S_0^2=S_1^2+S_2^2+S_3^2$, and for partially polarized light, $S_0^2<S_1^2+S_2^2+S_3^2$. The fully polarized states thus lie on a sphere in $(S_1,S_2,S_3)$ space, which is known the Poincar\'e sphere. The Poincar\'e sphere perfectly unifies the Stokes parameter and polarization ellipse pictures: Lines of constant $\Psi$ are longitudes on the sphere and lines of constant $\theta$ are the sphere's latitudes. The poles of the Poincar\'e sphere ($\theta = \pm \pi/4$) represent circularly polarized light of opposite handedness. 

\subsection{Optical fields as mappings}

 At every point of an $D_{parameter}$-dimensional parameter space, the optical field assigns a set of $D_{condition}$ real-valued numbers, such as the real and imaginary parts of its Cartesian fields. More formally, we can say that optical fields are maps from parameter space to condition space $F:S_{parameter}\mapsto S_{condition}$. $S_{parameter}$ is usually Cartesian space $S_{parameter}=\mathbb{R}^{D_{domain}}$ for simplicity, but both $S_{parameter}$ and $S_{condition}$ spaces just need to be differentiable manifolds; that is, they are spaces that locally resemble Euclidean space. Singularities are the positions in parameter space at which the $D_{condition}$ numbers are simultaneously zero; singular positions are the \textit{preimage} in parameter space of the \textit{origin} (all conditions zero) in condition space. Although this formalism may seem excessively convoluted when dealing with easily-visualised structures in scalar fields, it becomes essential when handling vector-valued fields with high dimensional $D_{domain}$ and $D_{condition}$ spaces, such as singularities in the polarization field of light.

There is a distinction to be made in the definition of the C loci topological invariant as compared to those cited in literature \cite{1983_Nye_PRSLA,2002_Freund_OptComm}. It is typically the accumulation of the azimuth $\Psi$ that is used for this invariant calculation, instead of the angle subtended by $\{S_1,S_2\}$, which is $2\Psi$. $\Psi$ assumes values from $[0,\pi)$ rather than $[0,2\pi)$, as a standard phase would, and accumulates half-integer and integer multiples of $2\pi$ on closed loops around the singularity. The reason is that the polarization azimuth does not have a pointing direction - it aligns with the semi-major axis of the polarization ellipse but remains invariant under a $\pi$ rotation, thus still referring to the same polarization ellipse. Polarization azimuths that differ by an integer multiple of $\pi$ are identical. Consequently, the topological invariant associated with accumulation of the polarization azimuth can take on both integer and half-integer values, whereas the topological invariant computed by accumulating $atan2(S_2,S_1)$ will only take integer values. In a 2D plane, generic C loci have three distinct topologies, named after their equivalent umbilical points: star $(I_C=-1/2)$, monstar $(I_C=+1/2)$, and lemon $(I_C=+1/2)$ \cite{1983_Nye_PRSLA}.

\subsection{Topological charge for C and V loci}

The topological charge for C loci is computed in a similar manner to phase singularities, since both singularities have a co-dimension of two. Consider a plane featuring a C polarization singularity at its center. At this singularity, the transverse polarization is strictly circular (or undefined), as that is the only condition that satisfies the condition set $\bm{E}_\perp \cdot \bm{E}_\perp =0$ or $\bm{E}\cdot\bm{E}=0$. Circular polarization means that the polarization azimuth is undefined. It was Freund who noted that the polarization azimuth $\Psi$ could be expressed as half the complex angle of the Stokes field $\sigma = S_1+iS_2$ (Equation \ref{eqn:cloci}), with $S_1$ and $S_2$ being two of the real-valued Stokes parameters \cite{2002_Freund_OptComm}. The Stokes field $\sigma$ is zero at the C locus itself $S_1=0, S_2=0$, thus the condition values are $\{S_1,S_2\}$. 
\begin{gather}
\arg(\sigma) = \arg(S_1+iS_2) = 2\Psi \label{eqn:cloci}
\end{gather}

The topological invariant associated with C loci can thus be computed by a closed loop integral on a plane around the singularity.
\begin{gather}
    l_{C} = \frac{1}{2\pi}\oint_S d^1\Omega = \frac{1}{2\pi} \oint_S \nabla atan2(S_2,S_1) \cdot d\bm{r}, l_C\in\mathbb{Z}
\end{gather}

Just as phase singularities with $D_{domain}=2$ are topologically protected when their winding numbers are $\pm 1$, a point singularity with co-dimension $D_{condition}$ and in a parameter space with $D_{domain}=D_{condition}$ is topologically protected when its topological charge (obtained from the degree of continuous mapping) is $\pm 1$. We can make this concrete by studying transverse V loci, which occur when the transverse complex polarization fields vanish $E_{x,y}=0$, a singularity with a co-dimension of four ($Re(E_x)=0,Im(E_x)=0,Re(E_y)=0,Im(E_y)=0$). To examine the topological protection of a point transverse V locus, we need to identify a four-dimensional parameter space. As described in Spaegele et al \cite{2023_Spaegele_Capasso_SciAdv}, such a parameter space is the three dimensions of Cartesian space and one synthetic dimension of incident wavelength: $(x,y,z,\lambda)$. Such singularities occur in parameter space when the fields map to the origin of condition space $\{Re(E_x),Im(E_x),Re(E_y),Im(E_y)\}$. To examine topological protection, we consider the mapping of the 4D ball surrounding the singularity in parameter space, which we will call $S$ to the 4D condition space:
\begin{gather}
    l_{4} = \frac{1}{2\pi^2}\oint_S d^3\Omega, l_4\in\mathbb{Z}
\end{gather}
$d^3\Omega$ is the differential solid angle for a 4D sphere (in condition space), which has a total solid angle of $A=2\pi^2$. When $l_4=\pm 1$, all possible values of $Re(E_{x,y}),Im(E_{x,y})$ occur in the immediate vicinity of the singularity (up to overall normalization), which means that all complex transverse polarization fields (including the global phase) exist in the 4D parameter space around that singularity. Such a transverse V locus is thus topologically protected against the perturbative influence of stray light of any transverse polarization. 
  
\subsection{Toraldo pupil synthesis}

How can Toraldo pupil synthesis produce diffraction lobes that are narrower than the angular resolution limit of $1.22\lambda/D$ as expressed in the Rayleigh criterion? The Rayleigh criterion is an angular resolution bound for open, unstructured circular apertures and does not set an angular resolution limit for structured apertures. Figure \ref{suppfig1} provides a visual explanation of Toraldo synthesis. The main goal of the method is to enforce zero intensities at particular diffraction angles in the far field. We begin by dividing a circular pupil into multiple concentric zones. We would like to solve for the transmission amplitude associated with each pupil zone so that the pupil will produce the desired diffraction pattern in the far field. This transmission amplitude is a real number for amplitude-only masks and can be complex for masks with phase control. Each concentric ring produces a Bessel function (of the first kind) in the far field under uniform illumination, and the total far-field diffraction pattern is the linear combination of these Bessel functions weighted by their respective (complex) amplitudes. Since there are $N$ degrees of freedom for $N$ concentric zones corresponding to their $N$ amplitudes $A_1,\ldots,A_N$, we can solve for $N$ independent constraints on the far-field diffraction pattern. In effect, we are looking for an \textit{interpolation} for a set of angle-intensity constraints in the far-field diffraction pattern over the $N$ individual ring diffraction patterns. The $N$ unknowns and $N$ constraints form a set of linear equations that can be solved numerically for $A_1,\ldots,A_N$. Schelkunoff and Di Francia's insight was that choosing these $N$ constraints to be \textit{zeros} of the diffracted field (red circles in Figure \ref{suppfig1}) could ``squeeze'' the main lobe of the angular pattern so that it becomes smaller than that of the unstructured aperture. Additional zeros can then be added to move the large sidelobes away from the squeezed main lobe. For long wavelengths in the radio regime, enough zero positions can even push the sidelobes into the evanescent field with diffraction angles greater than $\pm \pi/2$, ensuring that the sidelobes do not propagate into the far-field at all.

The central lobe in Toraldo synthesis can be made superoscillatory. The spatial frequencies in the far-field diffraction pattern come from the off-axis zones of the illuminated pupil: zones or rings that are further away from the optic axis produce Bessel functions with higher spatial frequencies. The maximum spatial frequency in the far-field of the pupil is determined by the diameter of the pupil; an infinitely thin ring with the same diameter as the pupil produces a Bessel function that has a central lobe full-width-at-half-maximum (FWHM) of $0.72\lambda/D$\cite{2007_Grosjean_Courjon_OptComm}. Thus, an engineered central lobe with a FWHM less than $0.72\lambda/D$ is superoscillatory.

The subdiffraction capabilities of Toraldo pupils come at a high cost: such supergain (super-directive in modern terminology) antennas with evanescent sidelobes require extremely high reactive currents and high current precision in the antenna elements, and these requirements become more stringent with higher gain. There is no theoretical upper limit to the gain of a supergain antenna; the limits come from the practicality of fabricating extremely precise antenna elements that can be strongly driven with precise currents\cite{1952_DiFrancia_IlNuovoCimento}.

\subsection{Singularities in the canonical superoscillation function}

It is sometimes unclear what role singularities play when the superoscillatory function itself does not contain any singularities. Consider the canonical superoscillation function, which is arguably the most studied superoscillatory function in literature\cite{2006_Berry_Popescu_JPhysA,2019_Berry_Hong_JOpt}:
\begin{gather}
f(x) = \left(\cos x + ia\sin x\right)^N, \quad x \in \mathbb{R}
\end{gather}

Where $N$ is a large even integer and $a>1$. The function's periodicity is $\pi$ and it is bandlimited with wavenumbers $|k|\leq N$. The function's intensity and phase profile are plotted in Figure \ref{suppfig2}a for $N=20$ and $a=10$. The function achieves very rapid phase variation near $x=0$ and large sidebands away from it. The behavior of the function near $x=0$ is plotted in Figure \ref{suppfig2}b. The first order approximation for $f(x)$ near $x=0$ is $e^{iaNx}$, thus it appears to have a large wavenumber $aN$ despite being bandlimited to $|k|\leq N$. $a>1$ can be considered the degree of superoscillation over the maximum frequency component. The function intensity $|f(x)|^2=[1 + (a^2 -1)\sin^2 x]^{2N}$ is strictly positive over the real line, hence there are no zeros or singularities where $f(x)=0$. 

When the canonical superoscillation function is plotted over the complex plane instead of just the real line (that is, allowing $x$ to be complex instead of purely real), the reason for the superoscillatory behavior of $f(x)$ immediately becomes clear. Figures \ref{suppfig2}(c-d) plot the intensity and phase profile of $f(x)$ over the complex plane for $x$, respectively. Although the real line does not pass through any singularities, there is a phase singularity located at a distance of $\sinh^{-1}[(a^2-1)^{-1/2}]$ above the real axis, producing a reduced function intensity along the real line around $Re(x)=0$. Importantly, the phase singularity is a \textit{vortex singularity} with a topological charge $l=N$. These complex plane images demonstrate the wider context around the 1D superoscillatory function, demonstrating that its rapid phase variation and low intensity near $Re(x)=0$ are simply due to its proximity to a charge $N$ vortex phase singularity located in the \textit{imaginary direction}! Increasing the degree of superoscillation $a$ decreases the distance $\sinh^{-1}[(a^2-1)^{-1/2}]$ to the phase singularity, thereby decreasing the minimum intensity and also increasing the maximum phase gradient along the real axis. Changing the value of the integer $N$ simply changes the number of $2\pi$ phase windings around the vortex core, thereby changing the number of $2\pi$ phase wrappings along the real line of the function. Even though the canonical superoscillation function does not pass through any singularities on its domain of the real line, its superoscillatory properties are still derived from its proximity to a phase singularity, albeit in the complex plane.

\newpage
\section{Supplementary Figures}
\setcounter{figure}{0}   

\begin{figure}[h!]
\centering
\includegraphics[width=10cm]{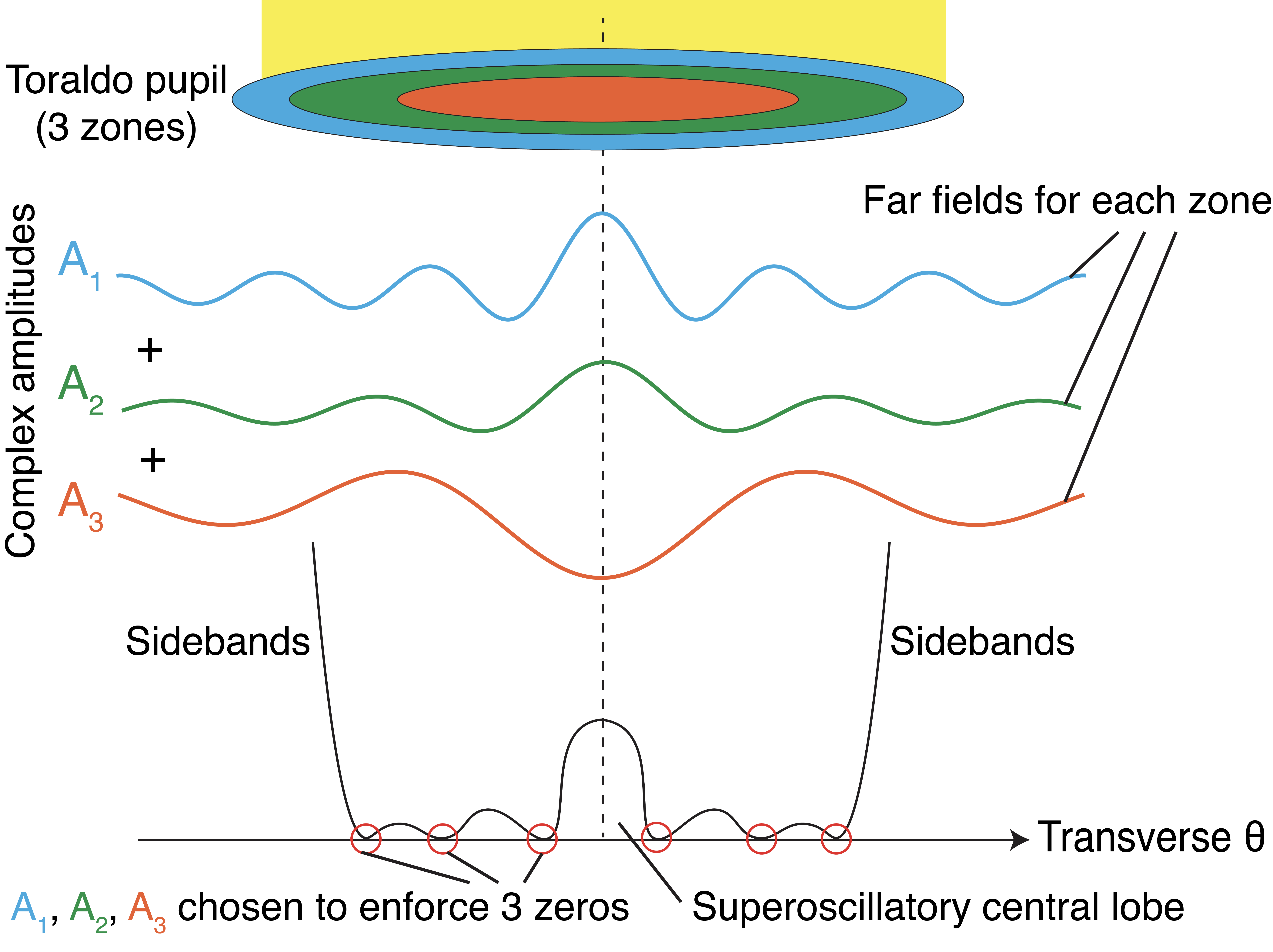}
\caption[Toraldo synthesis of superoscillatory spot in the far field]{Toraldo synthesis of superoscillatory spot in the far field, which can be used to construct super-directed antennae. A Toraldo pupil comprises various concentric zones. Each zone is associated with a complex or real-valued amplitude, depending if phase or amplitude masks are used. The far-field diffraction pattern can be written as the sum of the individual zone diffraction patterns, weighted by their respective amplitudes. By enforcing zeros in the far-field pattern, the central diffraction lobe can be squeezed until it is superoscillatory and has a characteristic width smaller than the diffraction limit. The cost is that large sidebands will form outside the superoscillatory region.
\label{suppfig1}}
\end{figure}

\begin{figure}[h!]
\centering
\includegraphics[width=12cm]{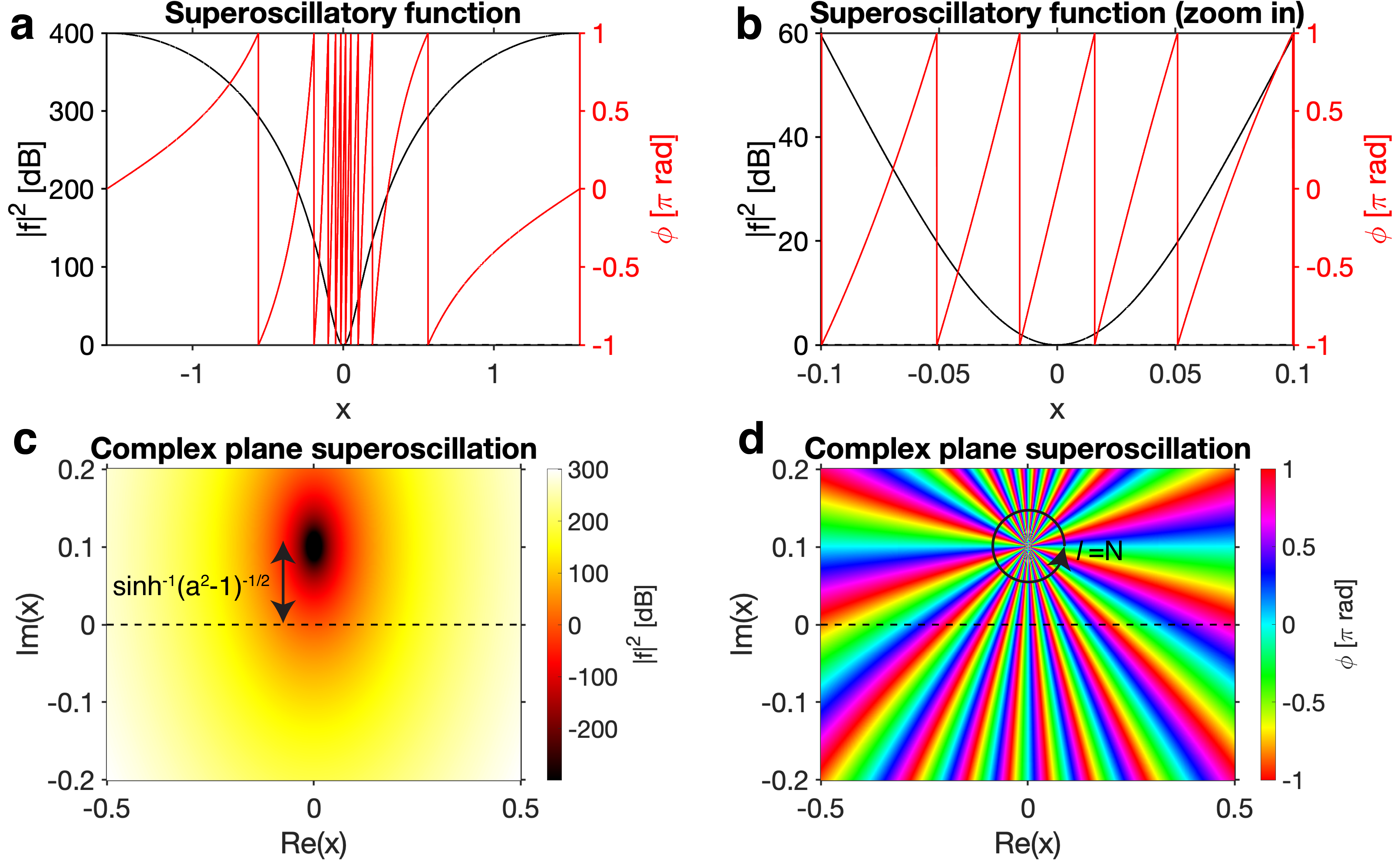}
\caption[Singularities of the canonical superoscillation function in the complex plane]{The canonical superoscillation function $f(x) = (\cos x + ia\sin x)^N$ is bandlimited to $|k|\leq N$ (shortest wavelength $2\pi/N$) and derives its superoscillatory behavior from nearby singularities in complex space. In this figure, $a = 10, N = 20$. \textbf{a} The canonical superoscillation function plotted on its periodic real domain of $x\in[-\pi/2,\pi/2]$ shows rapid phase accumulation behavior near $x=0$. \textbf{b} Zooming into the region near $x=0$, the function has a phase accumulation like $\exp(iNax)$, which is $a=10$ times more rapid than its spatial frequency bandlimit. The function does not vanish on the entire real line and so has no singularities on the real line. \textbf{c}-\textbf{d} The phase singularity responsible for the superoscillatory behavior is revealed when $x$ is allowed to be complex. The \textbf{c} intensity and \textbf{d} phase profiles of the complex plane superoscillation are plotted. A vortex singularity with topological charge $l=N$ is located at a distance of $\sinh^{-1}[(a^2-1)^{-1/2}]$ above the real axis. The larger the value of $a$, the closer the singularity is to the real axis, which increases the phase gradient along the real-axis.
\label{suppfig2}}
\end{figure}

\end{document}